\documentclass{pasj00}
\draft

\begin{document}
\SetRunningHead{Author(s) in page-head}{Running Head}
\Received{2006/08/12}
\Accepted{2006/10/02}

\title{Hard X-ray Detector (HXD) on Board  Suzaku}
 \author{Tadayuki \textsc{Takahashi}\altaffilmark{1,2},
  Keiichi \textsc{Abe}\altaffilmark{3},
  Manabu \textsc{Endo}\altaffilmark{4},
  Yasuhiko \textsc{Endo}\altaffilmark{3}$^\ast$,\\
  Yuuichiro \textsc{Ezoe}\altaffilmark{1},
  Yasushi \textsc{Fukazawa}\altaffilmark{5},
  Masahito  \textsc{Hamaya}\altaffilmark{6}
  Shinya \textsc{Hirakuri}\altaffilmark{2}$^\dag$,\\
  Soojing \textsc{Hong}\altaffilmark{3,7}$^\ddag$,
  Michihiro \textsc{Horii}\altaffilmark{6},
  Hokuto \textsc{Inoue}\altaffilmark{1,2}$^\S$,
  Naoki \textsc{Isobe}\altaffilmark{7},
  Takeshi \textsc{Itoh}\altaffilmark{2},\\
  Naoko \textsc{Iyomoto}\altaffilmark{2} $^\P$
  Tuneyoshi  \textsc{Kamae}\altaffilmark{8},
  Daisuke \textsc{Kasama}\altaffilmark{2}$^|$ 
  Jun \textsc{Kataoka}\altaffilmark{1,2}$^{\ast\ast}$\\
  Hiroshi \textsc{Kato}\altaffilmark{7},
  Madoka \textsc{Kawaharada}\altaffilmark{2},
  Naomi \textsc{Kawano}\altaffilmark{5}
  Kengo \textsc{Kawashima}\altaffilmark{5},\\
  Satoshi \textsc{Kawasoe}\altaffilmark{5},  Tetsuichi \textsc{Kishishita}\altaffilmark{1,2},
  Takao \textsc{Kitaguchi}\altaffilmark{2},
  Yoshihito \textsc{Kobayashi}\altaffilmark{1,2}$^{\dag\dag}$,\\
Motohide \textsc{Kokubun}\altaffilmark{2}, 
   Jun'ichi \textsc{Kotoku}\altaffilmark{2}$^{\ast\ast}$,
  Manabu \textsc{Kouda}\altaffilmark{1,2} $^{\ddag\ddag}$,
  Aya  \textsc{Kubota}\altaffilmark{7},\\
  Yoshikatsu \textsc{Kuroda}\altaffilmark{4},
  Greg \textsc{Madejski}\altaffilmark{8},
  Kazuo \textsc{Makishima}\altaffilmark{2,7},
  Kazunori \textsc{Masukawa}\altaffilmark{4},\\
  Yukari \textsc{Matsumoto}\altaffilmark{2}$^\dag$
  Takefumi \textsc{Mitani}\altaffilmark{1,2},
  Ryohei \textsc{Miyawaki}\altaffilmark{2},
  Tsunefumi \textsc{Mizuno}\altaffilmark{5},\\
  Kunishiro \textsc{Mori}\altaffilmark{9},
  Masanori \textsc{Mori}\altaffilmark{3,7}$^{\S\S}$,
  Mio \textsc{Murashima}\altaffilmark{2}$^{\P\P}$,
  Toshio \textsc{Murakami}\altaffilmark{10},\\
  Kazuhiro \textsc{Nakazawa}\altaffilmark{1},
  Hisako \textsc{Niko}\altaffilmark{2},
  Masaharu \textsc{Nomachi}\altaffilmark{11},
  Yuu \textsc{Okada}\altaffilmark{2}$^\dag$,\\
  Masanori \textsc{Ohno}\altaffilmark{5},
  Kousuke  \textsc{Oonuki}\altaffilmark{1,2},
  Naomi \textsc{Ota}\altaffilmark{7},
  Hideki \textsc{Ozawa}\altaffilmark{11},\\
  Goro \textsc{Sato}\altaffilmark{1,2} $^\P$,
  Shingo \textsc{Shinoda}\altaffilmark{12},
  Masahiko \textsc{Sugiho}\altaffilmark{2}$^{||}$ 
  Masaya \textsc{Suzuki}\altaffilmark{2}$^{\ast\ast\ast}$,\\
   Koji \textsc{Taguchi}\altaffilmark{6},
 Hiromitsu  \textsc{Takahashi}\altaffilmark{5},
  Isao  \textsc{Takahashi}\altaffilmark{2}$^{\dag\dag\dag}$\\
Shin'ichiro \textsc{Takeda}\altaffilmark{1,2},  Ken-ichi \textsc{Tamura}\altaffilmark{1},
  Takayuki \textsc{Tamura}\altaffilmark{1,2},\\
  Takaaki \textsc{Tanaka}\altaffilmark{1,2},
  Chiharu \textsc{Tanihata}\altaffilmark{1,2},
  Makoto \textsc{Tashiro}\altaffilmark{3},\\
  Yukikatsu \textsc{Terada}\altaffilmark{7},
  Shin'ya \textsc{Tominaga}\altaffilmark{5},
  Yasunobu  \textsc{Uchiyama}\altaffilmark{1,2},
  Shin \textsc{Watanabe}\altaffilmark{1,2},\\
  Kazutaka  \textsc{Yamaoka}\altaffilmark{13},
  Takayuki \textsc{Yanagida}\altaffilmark{2},
  and
  Daisuke \textsc{Yonetoku}\altaffilmark{10},
  }

  \altaffiltext{1}{Department of High Energy Astrophysics,
  Institute of Space and Astronautical Science (ISAS), \\
  Japan Aerospace Exploration
  Agency (JAXA), 3-1-1 Yoshinodai, Sagamihara, 229-8510}
  \email{takahasi@astro.isas.jaxa.jp}
    \altaffiltext{2}{Department of Physics, Graduate School of Science,
  	University of Tokyo, Hongo 7-3-1, Bunkyo, 113-0033}
\altaffiltext{3}{Department of Physics, Saitama University,
  	255 Shimo-Okubo, Sakura, Saitama, 338-8570}
  \altaffiltext{4}{Mitsubishi Heavy Industry, Co., Ltd.,
  	Nagoya Guidance and Propulsion Systems Works, \\
  	1200 Higashi Tanaka, Komaki, 485-8561}
  \altaffiltext{5}{Department of Physical Science, Hiroshima University,
  	1-3-1 Kagamiyama, Higashi-Hiroshima, 739-8526}
  \altaffiltext{6}{Meisei Electric, Co., Ltd.,
  	2223 Naganuma-cho, Isezaki, 372-8585}
  \altaffiltext{7}{The Institute of Physical and Chemical Research
  (RIKEN),
  	2-1 Hirosawa, Wako, 351-0198}
  \altaffiltext{8}{Stanford Linear Accelerator Center (SLAC),
  	2575 Sand Hill Road, Menlo Park, CA 94025, USA.}
  \altaffiltext{9}{Clear Pulse, Co., Ltd., 6-25-17 Chuuou, Ota, 143-0024}
  \altaffiltext{10}{Department of Physics, Kanazawa University,
  	Kakuma-machi, Kanazawa, 920-1192}
  \altaffiltext{11}{ Department of Physics, Osaka University,
 	1-1 Machikaneyama, Toyonaka, 560-0043}
  \altaffiltext{12}{ SSD Corp., 3-34-13-301 Togashira, Toride, 302-0034}
  \altaffiltext{13}{ Department of Physics and Mathematics
  	Aoyama Gakuin University,\\
  	5-10-1 Fuchinobe, Sagmihara, 229-8558}

\maketitle

\begin{abstract}
 The  Hard X-ray Detector (HXD) on  board Suzaku covers 
a wide energy range from 10 keV to 600 keV by combination of silicon PIN diodes and
GSO scintillators. The HXD is designed to achieve an extremely low in-orbit background
based on a combination of new techniques, including the concept of well-type active shield
counter. With an effective area of 142 cm$^{2}$ at 20 keV and 273 cm$^{2}$ at 150 keV, 
the background level at the sea level reached 
$\sim$ 1$\times$10$^{-5}$ cts s$^{-1}$ cm$^{-2}$ keV$^{-1}$ at 30 keV for the PIN diodes, and 
$\sim$ 2$\times$10$^{-5}$ cts s$^{-1}$ cm$^{-2}$ keV$^{-1}$ at 100 keV,
and $\sim$ 7$\times$10$^{-6}$ cts s$^{-1}$ cm$^{-2}$ keV$^{-1}$ at 200 keV for the
phoswich counter. Tight active shielding of the HXD results in a large
array of guard counters surrounding the main detector parts. These 
anti-coincidence counters, made of $\sim$4 cm thick BGO crystals,
have a large effective area for sub-MeV to MeV $\gamma$-rays. 
They work as an excellent $\gamma$-ray burst monitor with limited
angular resolution ( $\sim$ 5$^{\circ}$).  The on-board signal-processing system and the data
transmitted to the ground are also described.
\end{abstract}

\section{Introduction}

Suzaku is the fifth in a series of Japanese  X-ray astronomy satellite 
 with important US instrument contributions\citep{Mitsuda2006}. Its
scientific payload consists of two kinds of co-aligned 
instruments,  the X--ray Imaging Spectrometers (XIS) \citep{Koyama2006}
 and the Hard X--ray Detector (HXD). The HXD extends the bandpass of the observatory 
 to much higher energies with its 
10-600 keV  bandpass.
While the bandpass of previous Japanese 
X-ray satellites was primarily below $\sim$20 keV, where 
thermal emission predominates, the energy range of 10 - several 100 keV is where 
the radiation from the high-energy celestial sources is mainly non-thermal.  
There were several missions that covered this energy range,
but sensitive observations were very difficult,
since the signal from X-ray sources is 
much weaker than the detector background, 
and also because the fluctuations of such background are 
large at orbital environments in inclined low Earth orbits.

The main challenge of hard X-ray to gamma-ray spectroscopy is that the signal intensity 
from the source is usually weaker than detector background. The main causes of the
background include diffuse gamma rays, emission from radioactive nuclei due to 
activation of the detector, itself, and also cosmic rays such as protons and other heavy
ions. Besides, Compton down-scattering, dominant at higher energies,
 makes a proper X-ray event appear as background at a lower energy.
Based on the experience accumulated through a series of balloon 
experiments (\cite{Kamae92,Takahashi93}), the HXD is designed to minimize the
background level by utilizing two concepts: ``Well-type active shield'' and ``Compound-eye configuration''
and hence to achieve a higher sensitivity than any
previous instrument in the energy range between 10 keV and several 100 keV.  

Since Suzaku is the recovery mission to the original $Astro$-$E$, which was lost in the
launch accident in 2000, its basic design is the same as that of $Astro$-$E$.
We have designed and developed the HXD basically in the same manner as the
previous HXD  on board the lost $Astro$-$E$ (\cite{Kamae96,Tashiro02}). However, we   employed a limited range of improvements,
particularly concerning the sensor and analog electronics. The HXD fabrication was
carried out involving detailed qualification at every step, with a special emphasis 
on the verification of the newly introduced improvements.
This paper describes the design of the HXD system and its ground performance. An 
accompanying paper (\cite{Kokubun2006}; hereafter Paper II) considers the in-orbit performance  of the HXD. 
Early reports on the HXD system and ground calibration can be found in previous 
publications (\cite{Kokubun03,Kawaharada04,Terada04,Fukazawa06,Yamaoka06} and references therein.)

\begin{figure}[ht]
\centering
\FigureFile(80mm,50mm){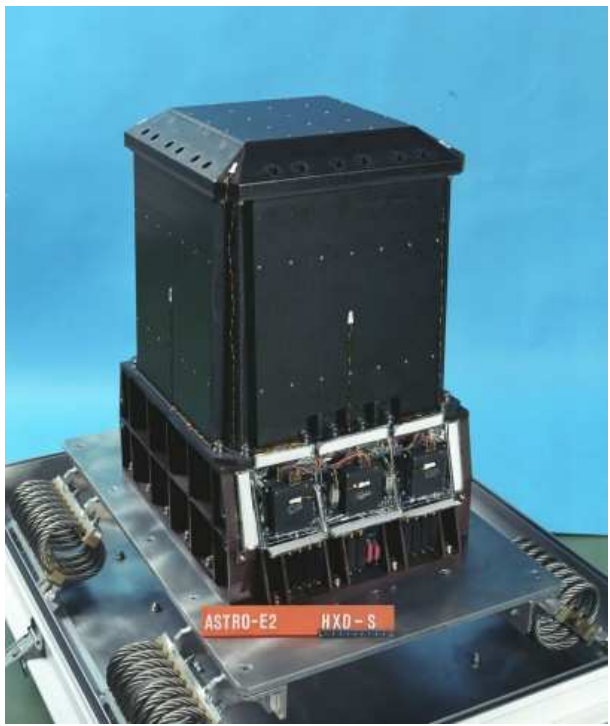}
\caption{Hard X--ray Detector before installation.  \label{Fig:hxdpic}}
\end{figure}

\begin{figure}
\centerline{\FigureFile(100mm,50mm){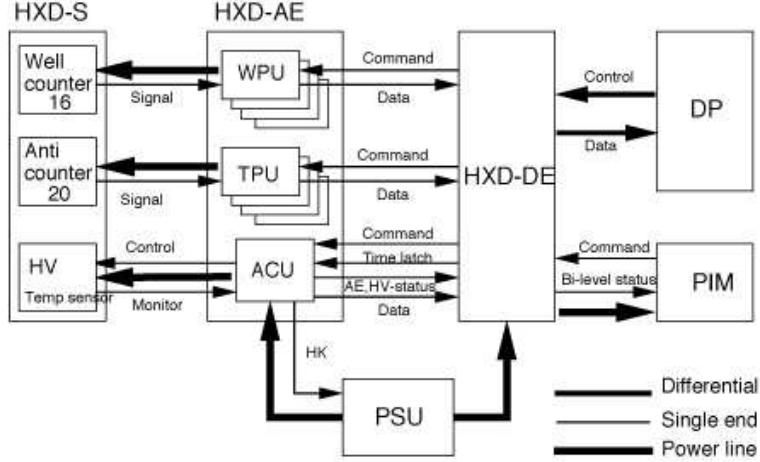}}
\caption{Block diagram of the HXD system and signal flow
between components.}
\label{Fig:Blockdiagram}
\end{figure}

\begin{figure}[ht]
\centering
\FigureFile(100mm,50mm){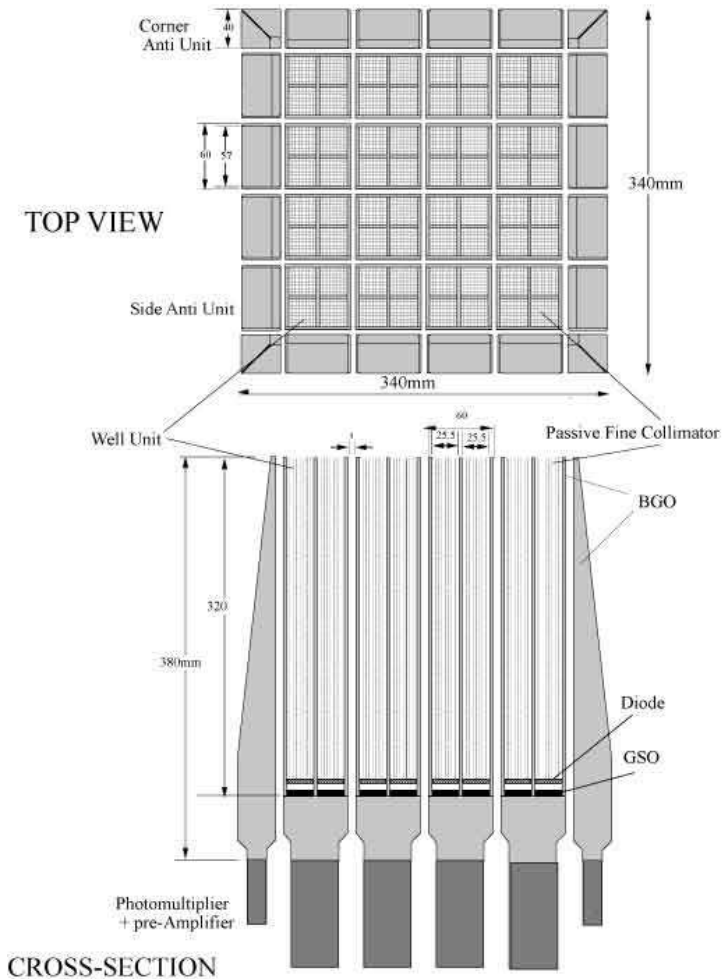}
\caption{Schematic drawing of HXD-S. It consists of 16 well-counter
units and 20 anti-counter units. \label{Fig:hxdall}}
\end{figure}

\section{The HXD System}

\begin{figure}[ht]
\centerline{\FigureFile(120mm,50mm){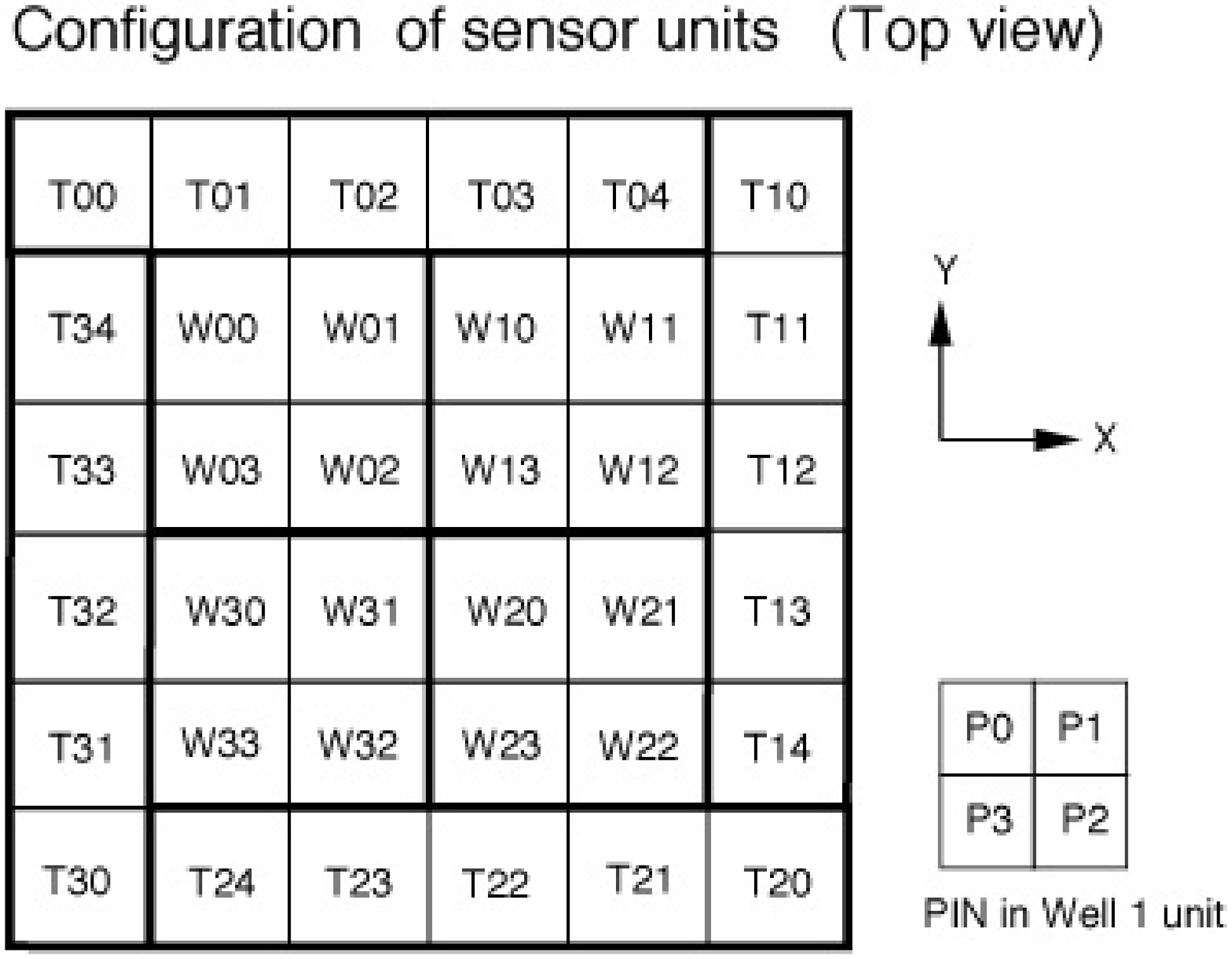}}
\caption{Numbering of the well and anti counter units when HXD-S
is viewed from the top. There are 16 well-counter units from W00 to W33
and 20 anti-counter units from T00 to T34. Y direction corresponds to 
the direction toward Sun when the HXD is mounted to the satellite.}
\label{Fig:sensor-number}
\end{figure}

The HXD is a non-imaging and collimated instrument. The
system is composed of the sensor part (HXD sensor or HXD-S), an analog electronic system
(HXD-AE), and a digital electronics system (HXD-DE). 
A photograph of HXD-S is shown in figure \ref{Fig:hxdpic}. 
As described in the block diagram (figure \ref{Fig:Blockdiagram}), HXD-AE receives   analog
signals from HXD-S, selects the proper events and sends their digitized
data  to HXD-DE.
 HXD-DE then formats the data packets to send them to the ground via the satellite 
data processor (DP). Commands to the HXD are received by the peripheral interface module (PIM)
 from the satellite data 
handling unit, and transmitted to HXD-DE. All power is supplied from a power-supply unit called HXD-PSU.
The operating temperature of HXD-S is designed to be  $-20^{\rm o}$C.
The characteristics of the HXD are summarized in table \ref{Table:capab}.

\begin{table*}
\caption{Characteristics of  the HXD system on board Suzaku \label{Table:capab}}
\begin{tabular}{l|ll} 
\hline \hline
HXD & Field of View & $4.5^{\circ}\times4.5^{\circ}$ ($\gtrsim 100$\,keV) \\
    & Field of View & $34' \times 34'$\ ($\lesssim 100$ keV) \\
    & Bandpass & 10 -- 600 keV \\
    & -- PIN   & 10 -- 70 keV \\
    & -- GSO   & 40 -- 600 keV \\
    & Energy Resolution (PIN) & $\sim 3.0$\,keV (FWHM) \\
    & Energy Resolution (GSO) & $7.6 / \sqrt{E_{MeV}}$ \% (FWHM) \\
    & Effective area & $\sim 160$\,cm$^2$\ at 20 keV, $\sim 260$\,cm$^2$\ at 100 keV\\
    & Time Resolution & 61 $\mu$s or 31 $\mu$s\\  \hline
HXD-WAM$^{a}$ & Field of View & 2$\pi$ (non-pointing) \\ 
        & Bandpass      & 50 keV -- 5 MeV \\ 
        & Effective Area & 800 cm$^2$\ at 100 keV / 400 cm$^2$\ at 1 MeV\\
        & Time Resolution & 15.625 ms or 31.25 ms for GRB, 1 s for All-Sky-Monitor\\ \hline
\end{tabular}

\vspace{1mm}

a: Shield counters of the HXD sensor part act as a wide-band all-sky monitor (WAM).
\end{table*}

\section{HXD Sensor (HXD-S)}

Since the background level sets the sensitivity limit in the hard X-ray band, the HXD
 is designed  to minimize the background by its improved 
phoswich (acronym for PHOSphor sandWICH) configuration for the energy region above 40 keV
and the adoption of newly-developed 
thick silicon PIN diodes  for the energies below $\sim$ 70 keV.
Our detector ensures a low background though
the following techniques.
\begin{enumerate}
\item Well-type active shield:

 In phoswich  counters, 
two crystals with different decay times are used for the detection part (faster decay 
time) and the shielding part (slower decay time), and both signals are extracted
by a single photomultiplier. The improvement is that the shield is shaped in a well,
so that it also acts as an active collimator (well-type active shield).
This narrows the field of view (FOV) of the phoswich counter  without 
any additional passive material,
and results in the main detection part 
having an active shield of almost 4$\pi$ of its surrounding (well-type phoswich counter).
In the HXD, the well-type shield provides very efficient shielding for the the PIN diode,
which is read out independently.

\item Compound eye configuration:

HXD is modular, that is, consisting of a number of units.  Each 
well-type phoswich counter unit has a simple shape and operates at a
modest count rate by itself. In the HXD, we increase the photon collecting area 
by placing individual units in a  matrix.  In this configuration, 
each unit also becomes an active shield for adjacent units 
(compound eye configuration). It is also useful to reduce the
possible dead time if  parallel processing of each unit
could be implemented.  For additional shielding ofr the 
outer most units,  thick anti-coincidence counters are placed 
surrounding the well units.
\end{enumerate}

The main detection part of the phoswich counter in HXD-S is a Gadolinium 
silicate crystal (GSO; Gd$_{2}$SiO$_{5}$(Ce)) buried deep in the bottom of a
well-shape bismuth germanate crystal (BGO; Bi$_{4}$Ge$_{3}$O$_{12}$)
(hereafter we refer it as well-counter unit).
As shown in the schematic drawing  (figure \ref{Fig:hxdall}),
the HXD-S consists of 16 phoswich counters with 4 silicon PIN diodes in
each, and 20 surrounding BGO anti-coincidence shield counters (anti-counter unit). 
All of the 16 well-counter units and 20 anti-counter units work 
 independently. The numbering of the
well-counters and anti-counters, which is frequently referred to in various technical
documents,  is shown in figure \ref{Fig:sensor-number}

During  operation of the HXD, instead of rocking observation, 
we plan to perform background subtraction 
by modeling the background spectrum,
as was done in the LAC detector in Ginga satellite \citep{Hayashida89}.
Thus the sensitivity will depend on the accuracy of the background 
modeling. If the background flux can be kept low,
any systematic error of the modeling will  have little effect on the spectrum 
 of sources with a given flux.  Since radioactive contamination
in the detection part contributes to the background directly,
  special care was taken to select materials for the components used in the HXD sensor.

\subsection{Well-counter units}

\begin{figure}[htbp]
\begin{center}
\FigureFile(100mm,50mm)
{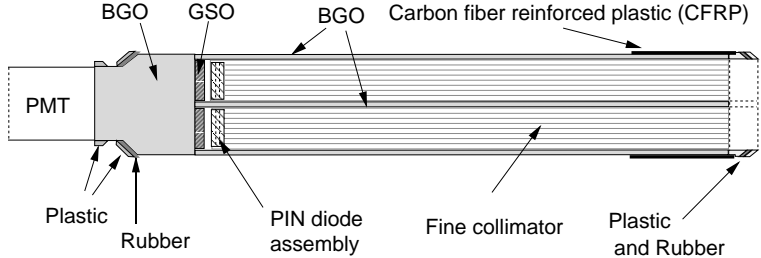}
\caption{
A schematic drawing of the well-counter unit. The  silicon PIN 
diode with dimensions of 21.5 mm
$\times$ 21.5 mm $\times$ 2 mm is placed in the deep
well just above the GSO scintillator. The 24 mm $\times$ 24 mm GSO scintillator
with a thickness of 5 mm
is glued to the bottom part of the well-type BGO shield.
\label{Fig:hxd-well}
}
\end{center}
\end{figure}

In the HXD, each well-counter unit consists of five components (figure \ref{Fig:hxd-well})  
crystals, PIN diodes, fine collimators, a photomultiplier,
and front-end electronics. The mass of each unit is 4.63 kg.
The crystals
 in the well-counter unit are four 5~mm-thick GSO scintillators with dimensions of
24 mm $\times$ 24 mm,
and BGO scintillators for active shields.  A photomultiplier, HAMAMATSU R6231-07, is attached to the 
counter unit to collect scintillation light from both GSO and BGO scintillators.
The dynode signal from the photomultiplier is first processed by the preamplifier
mounted in the housing of HXD-S,
In the well-counter unit, the BGO scintillator is further divided into two pieces;
a block-like section called ``BGO bottom piece'', and a well-shaped
long section called ``BGO top piece''. The BGO top piece is not a
simple ``well'', but has cross-shaped inner BGO plates of 3~mm thickness to 
divide the ``well'' into four narrow collimators.  
The four GSOs are placed on the BGO bottom piece and surrounded by  four  long  square 
tubes of the BGO top piece.
The BGO top piece actively restricts the field of view to $4^{\circ}.6 \times 4^{\circ}.6$.
By thus collimating the field of view, and rejecting those events that deposit all, or part,
of energy
in BGO, the well-Counter unit attains high sensitivity.

In addition to their large stopping power, the reason
for having  two different types of crystals for the sensors
and the shields is 
their very different rise/decay times: of $\sim 706$ ns for BGO, and
$\sim 122$ ns for GSO, at a working temperature of $-20^{\rm o}$C \citep{Kamae92}.
This allows an easy discrimination of the sensor signal from the shield
signals, where a single photomultiplier  can discriminate between the two types
of scintillators in which an event may have occurred.  Any particle
events or Compton events that are registered by both the BGO and GSO
can be rejected by this phoswich technique.
The GSO/BGO phoswich has several advantages over more conventional (e.g., NaI/CsI)
combinations. Firstly, both GSO and BGO features a higher  stopping power for 
$\gamma$-rays.  Secondly, they have
 fast decay times suitable for a fast signal processing, and
a large difference in the decay times between the two components, which makes
 pulse-shape discrimination easier. 
Furthermore, the number of long-lived
$\gamma$-rays due to activations caused by charged particles irradiating the
detector is smaller, compared with other high-light-yield scintillation materials
such as YAP and NaI(Tl), if we normalize by the absorption thickness in the
hard X-ray energy range.
In table~\ref{Table:hxd-scinti}, we summarize the properties of
the GSO and the BGO scintillators,
together with those of NaI and CsI for a  reference. In the HXD, we use a GSO
crystal made by Hitachi Chemical  
\footnote{http://www.hitachi-chem.co.jp/english/index.html}
with
highly purified materials supplied by Shin-etsu Chemical Co Ltd. \footnote{http://www.shinetsu.co.jp/e/index.shtml}
in order to reduce any possible intrinsic background in the material.
This purification is found to be very effective to increase the light output
of the GSO scintillator \citep{Kamae02}

\begin{table*}[htb]
\begin{center}
\caption
{Properties of the GSO and BGO scintillators, together with those of NaI(Tl) and CsI(Tl) scintillators. }
\label{Table:hxd-scinti}
\begin{tabular}{|c | c  c  c  c | }
\hline
                                & GSO(Ce)               & BGO                      & NaI(Tl) & CsI(Tl) \\
                                & Gd$_{2}$SiO$_{5}$(Ce) & Bi$_{4}$Ge$_{3}$O$_{12}$ & NaI     & CsI     \\
\hline
Effective decay time (ns)@$20^{\circ}$C  & 86                    & 353                      & 230     & 1000   \\
Effective decay time (ns)@$-20^{\circ}$C  & 122                   & 706                      &      &    \\
Effective atomic number         & 59                    & 74                       & 50      & 54      \\
Density (g/cm$^{3}$)            & 6.7                   & 7.1                      & 3.7     & 4.5     \\
Radiation length (cm)           & 1.4                   & 1.2                      & 2.6     & 1.9     \\
Peak emission  (nm)             & 430                   & 480                      & 410     & 565     \\
Light yield (photons/MeV)$^{*}$       & $\sim$ 10000          & $\sim$ 4000              & 38000   & 32000   \\
Index of refraction (at $\lambda_{em}$)          & 1.9                   & 2.15                     & 1.85    & 1.80    \\
Hygroscopicity                  & none                  & none                     & yes     & little   \\
\hline
\end{tabular}
\end{center}

$^{*}$ Light yield measured with a bi-alkali photomultiplier.
\end{table*}

\begin{figure}
\centerline{\FigureFile(100mm,50mm){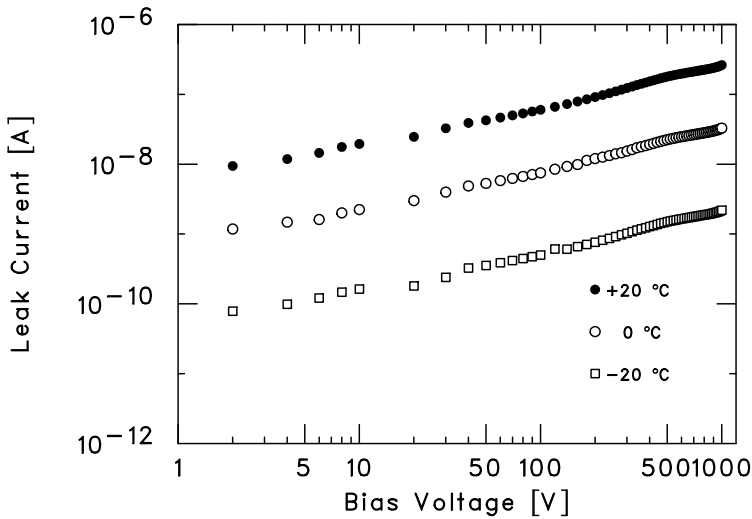}}
\caption{Typical leakage current of a 2 mm-thick silicon PIN diode developed for the
HXD as a function of the bias voltage. Data obtained at three different temperatures are shown. }
\label{Fig:PIN-IV}
\end{figure}

Since the  GSO  part covers an energy range of $\sim$ 40--600 keV, 
another detector element is needed to assure  coverage over the energy range 
down to that of the XIS  ($< 12$ keV).  The low energy response of the HXD is 
provided by 2 mm thick PIN
silicon diodes, each placed in front of a GSO crystal to form a PIN-GSO
pair as the detection part of the well-counter unit.  The diodes
absorb X--rays with energies below $\sim 70$ keV, but gradually become
transparent to harder X--rays, which in turn, reach and are registered
by,  the GSO detectors.  Such PIN diodes were specially developed jointly with 
Hamamatsu Photonics K.K. \footnote{http://jp.hamamatsu.com/en/index.html} to have 2 mm thickness \citep{Ota99}.  
These thick PIN diodes were manufactured from  highly purified silicon wafers, and
can be fully depleted at  a bias voltage of $\sim500$ V due to  their high 
resistivity (20--30 k$\Omega$ cm). In order to lower the leakage current
and attain stable operation, specially designed guard rings are built in the PIN diodes.
The geometrical area of the PIN diode, including the guard ring, is 21.5 mm
$\times$ 21.5 mm.
Figure \ref{Fig:PIN-IV} show  current-voltages
curves obtained from a typical PIN diode taken at   $-20^{\rm o}$C,  $0^{\rm o}$C and
$20^{\rm o}$C. 
The PIN diodes in the HXD have a leakage current of less than 
2.2 nA at 500 V and@ $-20^{\rm o}$C, 
and do not show any breakdown even under the bias voltage of 1000 V.  
In order to minimize the background gamma-ray contamination for the PIN,
we  selected low-background materiasl for the cables and the package. 
The adoption of a ceramic with a purity higher than 96 \% for the PIN package 
was required to avoid any continuum gamma-ray background due to
$\beta$-decay electrons from potassium in the material.

\begin{figure}
\centerline{\FigureFile(100mm,60mm){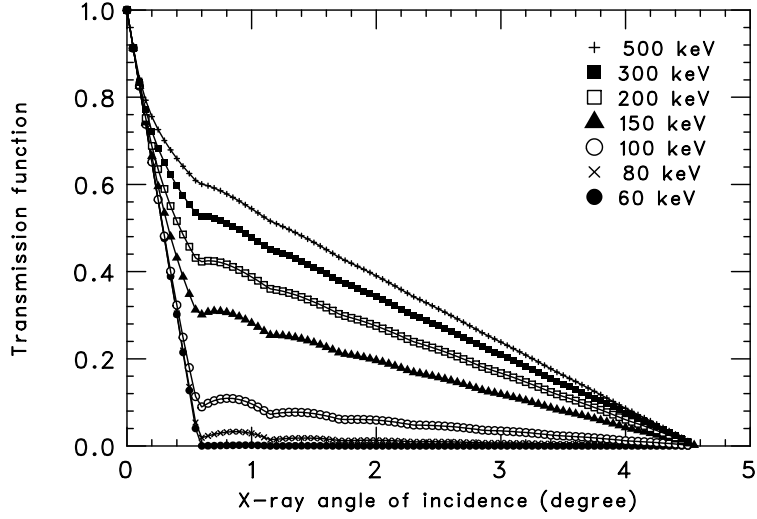}}
\caption{Angular transmission function of the fine collimator calculated at an azimuth angle of 0 degree. The 0 degree azimuth angle is defined along the positive x-axis in figure \ref{Fig:sensor-number} }
\label{Fig:calc-angres}
\end{figure}

\begin{figure}
\centerline{\FigureFile(100mm,60mm){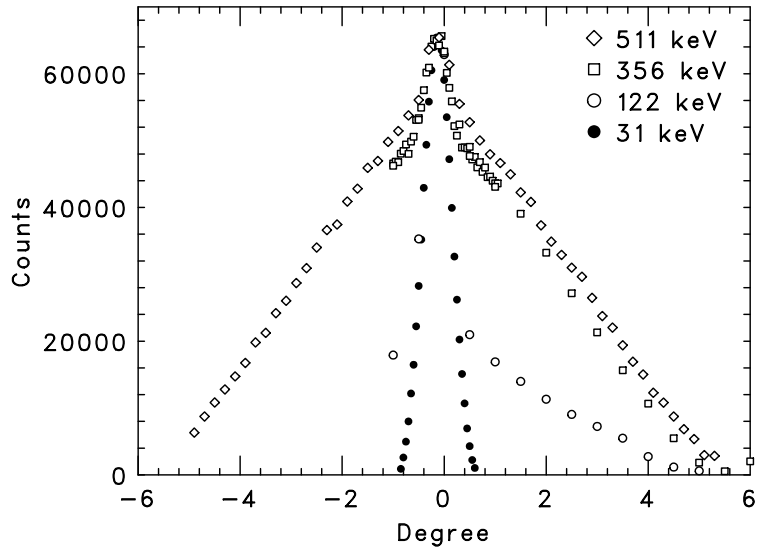}}
\caption{Angular responses of well-counter unit, measured with radio isotope sources (511 keV from $^{22}$Na,
356 keV from $^{133}$Ba,  122 keV from $^{152}$Eu, 31 keV from $^{133}$Ba) placed 280 cm away. The response is consistent with the calculation when the finite distance from the sources to the unit is taken into account.}
\label{Fig:measure-angres}
\end{figure}

In the energy range of the PIN diode, the dominant background 
component is the cosmic X-ray background. Therefore, having
a narrow FOV is the most effective method to reduce the
background contamination.
For this purpose, passive shields called ``fine collimators''
are inserted in the BGO well-type collimator  above the PIN diodes.
The fine collimator is made of 50 $\mu$m thick phosphor bronze sheets
arranged to form a square array of 
8 $\times$ 8 square channels each  of 3 mm width  and 300 mm length.
Both the BGO collimator and the fine collimator define the FOV
 of the well-counter unit. Because of the finite thickness, the FOV changes 
 with the photon energy.  Below
$\sim 100$ keV, the passive fine collimators define a $34' \times 34'$
full-width half-maximum (FWHM) square FOV. Above $\sim 100$ keV, the fine collimators become transparent and the
BGO active collimator defines a 4.5$^{\rm o}\times$ 4.5 $^{\rm o}$
FWHM square opening.  Figure \ref{Fig:calc-angres} shows the calculation of 
 the energy dependence of the angular response. 
 The collimator becomes transparent above $\sim$ 100 keV,
approaching the larger FOV of
$4.5^{\circ} \times 4.5^{\circ}$ (FWHM), defined by the BGO well.
The angular response measured with $\gamma$-ray lines from radio isotope sources
 located at limited distance is consistent  with the calculation (figure \ref{Fig:measure-angres}).

\begin{figure}
\centerline{\FigureFile(120mm,50mm){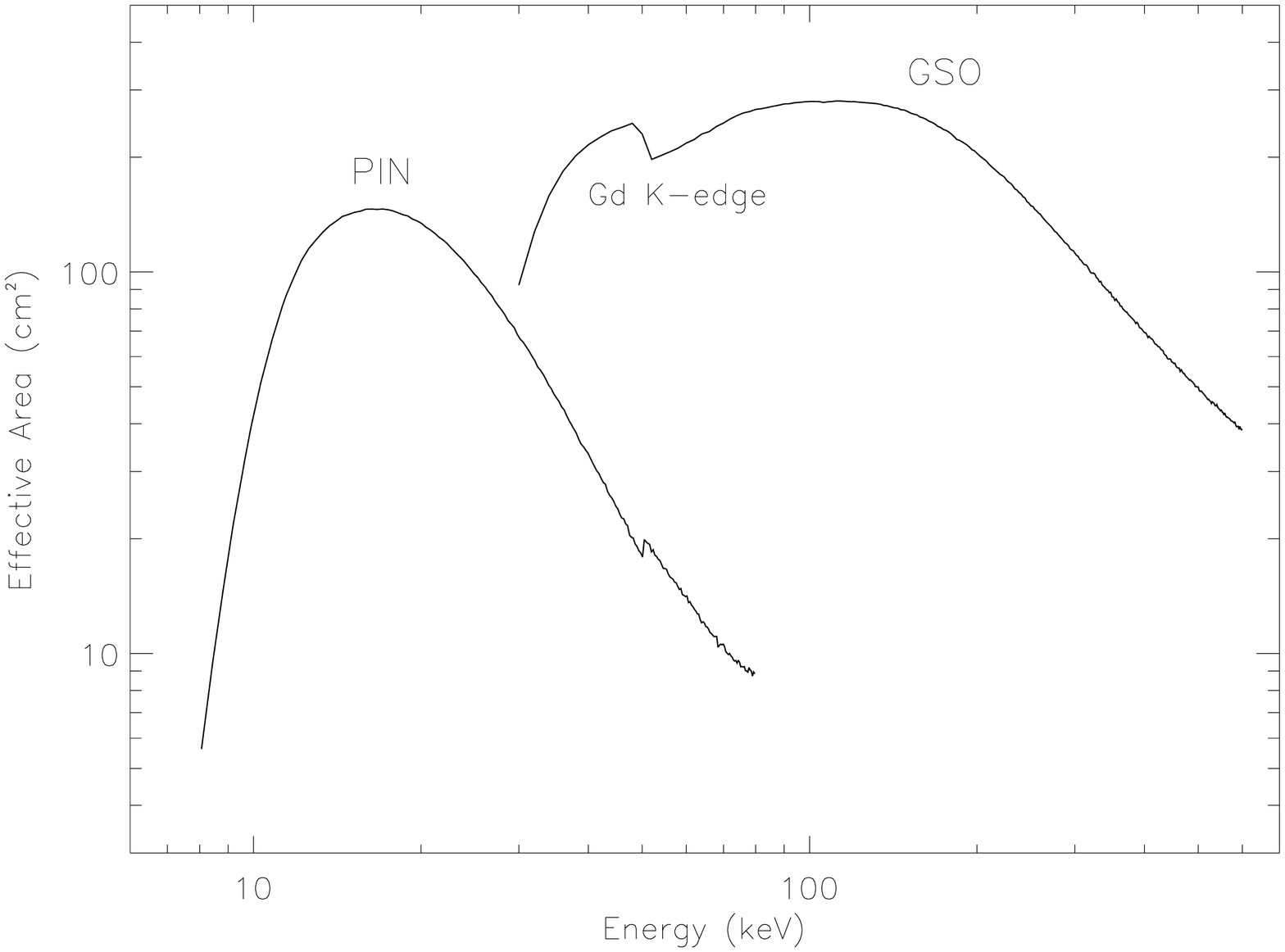}}
\caption{Total effective area of the HXD detectors, PIN and GSO, 
as a function of energy. Photon absorption by materials
in front of the device is taken into account.}
\label{Fig:hxd-eff}
\end{figure}

A Well-counter unit is completed by assembling all of the above and several
additional components, and 
further incorporating several additional attachments. 
In order to attain a high light-collection efficiency through the better light reflection 
at the surface, we  painted the BGO pieces with BaSO$_{4}$ and
wrapped them with a Gore-Tex sheet. Similarly, the top and side surfaces of each GSO was
wrapped by Gore-Tex sheets for better light reflection.
The mechanical support structures at the top and   bottom protect 
each well-counter unit  from any  vibration and shock
during the launch. The structures are made of
visco-elastic rubber, plastic adapters, and carbon fiber reinforced plastics (CFRP)
 plates. A detailed
description of the construction of the units is provided in Nakazawa et al. (1999).
Figure~\ref{Fig:hxd-eff} shows the effective area of 16 Well-counter units,
which was based on the photo-peak efficiency calculated by Monte Carlo simulations using the EGS4 code (\cite{Nelson85}).
 The total geometrical area of the 
PIN diodes is 160 cm$^2$, while that of the GSO crystals is 350 cm$^2$.
It is clear that the overlapping energy region between GSO and PIN, $\sim$ 40--70 keV, is 
well covered.

\subsection{Anti-counter units}

\begin{figure}[h]
   \begin{center}
   \begin{tabular}{c}
  \FigureFile(120mm,50mm)
{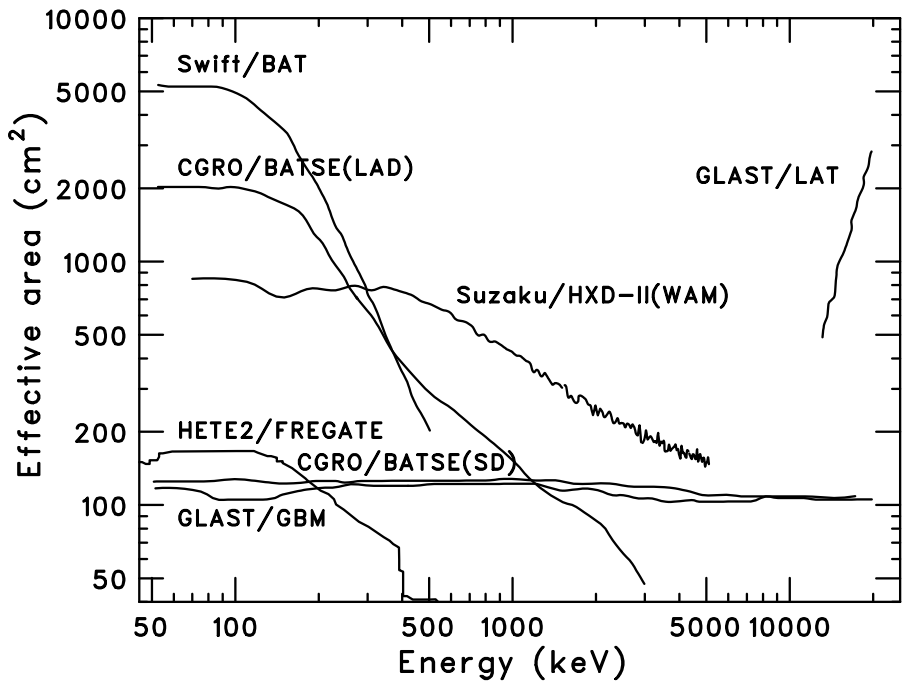}
   \end{tabular}
   \end{center}
   \vspace{-0.5cm}
   \caption{Comparison of the effective area of the anti-counter units per side
   (excluding the ``Corner Units'')
    and those of other GRB detectors. }
   \label{fig:anti_area}
   \end{figure}

For additional  shielding of the outermost well-counter
units, 20 anti-counter units, each made of   thick (average 2.6 cm) BGO pillars,
surround a well-counter matrix. This reduces the cosmic proton 
flux on the PINs and the GSOs by an order of magnitude. It also 
serves to reduce Compton scattered events as well as nuclear activation background events.  Each
unit is viewed by a  photomultiplier (HAMAMATSU R3998-01). The prime function
of the anti-counter units is to provide hit-pattern information to the Well-counter units.
The anti-counter units have a wedge-like structure for saving their mass while
obtaining the same path length of the BGO scintillator viewed from the PIN and the
GSO in the well-counter units. The mass of each unit is 4.12 kg for 16 units except
for the corner units with a mass of 2.72 kg.
The geometrical area of the anti-counter wall is as large as $\sim$ 800 cm$^{2}$
per side, and its effective area is $\sim$ 400 cm$^{2}$ even for 1 MeV $\gamma$-rays.
Therefore, they are also utilized as an excellent $\gamma$-ray burst detector in the energy
range between 50 keV and 5 MeV.
According to a MonteCarlo simulation, 
if a gamma-ray burst occurs with an intensity of 1 cnts/s/cm$^{2}$ in 50$-$300 keV, 
the spectrum with parameters of $\alpha = -1.0, \beta=-2.3$ and $E_{0}=250$ keV 
in the Band function (\cite{Band93}), and a duration of 20 sec, 
is captured by two side faces of the anti-counter units, we can determine its one 
dimensional position with an accuracy of $\sim$ 5$^{\circ}$, by comparing signal
counts from the four sides. 
As shown in figure~\ref{fig:anti_area}, the anti-counter units have the 
largest effective area compared with other current or future $\gamma$-ray burst (GRB) detectors,
including the GLAST/GBM and the Swift/BAT in the MeV range. Above 300 keV, the HXD  has
the largest effective area. The detailed description of the WAM and its in-flight performance
is described in Yamaoka et al. (2006).

\subsection{HXD Housing}

The housing structure for HXD-S is designed to hold   36 detector units
in HXD-S.
The function of the housing is to protect the units mechanically,
to minimize any temperature gradient among the units,
and to reduce electrical noise.
These requirements should be realized within the limited space and mass
allowed to the instrument. For this purpose, CFRP and Magnesium alloy are
adopted for the material to be used.

The  HXD-S housing  is made of three components:
the top lid and the center body part made of CFRP,
and the bottom structure made of Magnesium-alloy.
CFRP is selected not only for its strength and light mass;  
it also has a lower thermal expansion compared with the BGO crystal,
and will not give any  stress to the detectors once HXD-S is cooled to the operating
temperature of  $-20^{\rm o}$C in orbit.
The top lid has 64 square openings matching the FOV of the well units;
another thin ($400 \mu$m) plate of CFRP is used to cover them
for optical light blocking.
The total mass of the housing is 27.3 kg.

To achieve an operating temperature of $-20\pm5^{\rm o}$C, HXD-S is 
thermally decoupled from environment by multi-layer insulator (MLI),
and mounted 
on a cold-plate, which  is attached to the lower deck of the spacecraft.
The cold-plate is kept at low temperature via radiation cooling, and is also
equipped with a heater to control its temperature. 
The heat generated in the sensor part is $\sim 17$ W, half of which is from the
high voltage (HV)  units that are attached to the bottom structure of the housing.
Therefore the HV units are placed outside of the MLI,
thermally insulated from the main body of the  sensor,
and radiatively coupled to the spacecraft.
The remaining power is dissipated in the pre-amplifiers and PMT bleeders.
Except for the locations of the HV units and other heat sources,
 HXD-S is designed to achieve a rather uniform
temperature within a few degree.
The thermal design of HXD-S was verified in a thermal-vacuum
test using the spacecraft thermal model. Further details on the mechanical structure
 are described in Nakazawa et al. (1999).

\section{Onboard Data Processing System}
The electronic system of the HXD plays a crucial role 
in obtaining the best performance of the instrument. 
The total number of signal channels extracted from HXD-S 
to be processed is 116 (96 channels for the well units 
and 20 channels for the anti-counter units). 
A robust and stable electronic system is required to process
data from these channels under nominal constraints for a space experiment,
including in particular a severe power limitation.
Furthermore, the real time performance of the system is
critical to minimize the dead time caused by triggers issued randomly
from these channels. 
The onboard data processing system of the HXD
is thus carefully designed \citep{Takahashi98,Tanihata99},
incorporating two major system components,
namely an analog processing system, called HXD-AE (analog electronics), 
and a CPU-based processing system, called HXD-DE
(figure \ref{Fig:Blockdiagram}).
Here we briefly describe HXD-AE, HXD-DE, and the output data.

\subsection{Analog Electronics (HXD-AE)}

HXD-AE consists of one control board called Analog Control Unit (ACU) and eight signal processing boards. 
The latter is broken into two types, 
four well processing unit (WPU) boards for the Well counter units, 
and four transient processing unit (TPU) boards for the anti-counter units. 
While the well counter units acquire   data from target sources, 
the anti-counter units have an additional function of  
monitoring transient sources  (including gamma-ray bursts),
in addition to its basic role as active shields.The ACU is provided power from HXD-PSU,
and controls the power lines  to the other electronic boards in HXD-AE
as well as those to HXD-S. 
It also handles house keeping (HK) data, 
such as  high voltage values 
and temperatures of the scintillators and photomultiplier tubes.

Each WPU and TPU, respectively, handles four well units and five anti-counter units.   
All WPU/TPU and ACU boards are connected to the backplane of the AE housing. 
In order to reject the Compton events 
that occur in the compound eye configuration of the HXD,
it is important for each unit 
 to be informed of  the presence or absence of hits in the other 35 units. 
This information, called hit pattern, is shared by all
electronics boards through a ``hit-pattern bus'', which runs through the backplane. Besides the hit pattern information, other lines, 
such as power-lines and clock-lines also run through the backplane.

\begin{figure}
\centering
\FigureFile(100mm,50mm){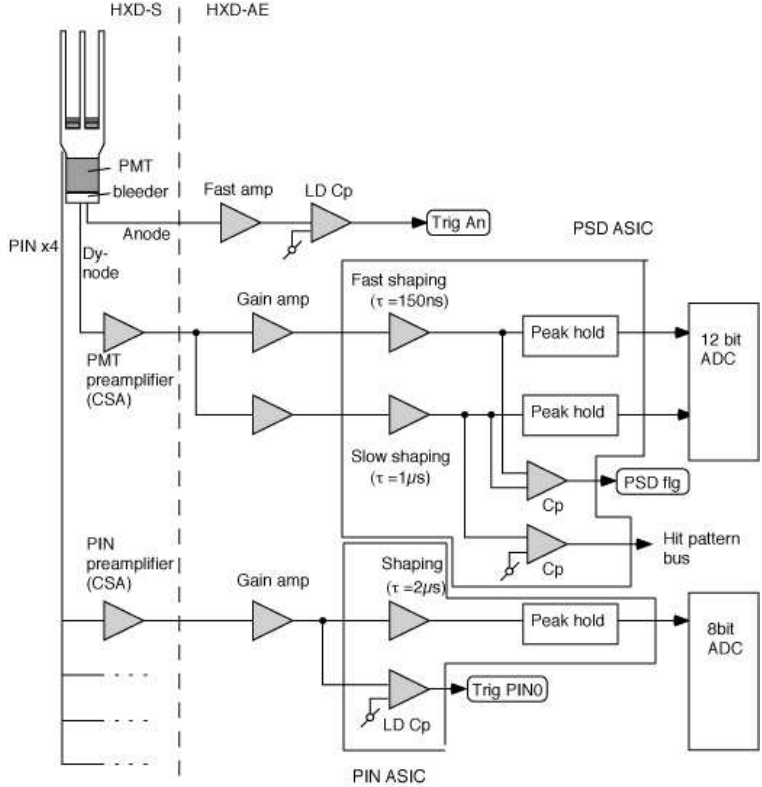}
\caption{Signal flow  from a  well counter unit   to a WPU board. 
	A pair of anode and dynode signals   from the bleeder and
	four PIN diode signals are processed in one block.
	 A WPU contains four of these blocks, and handles signals from four well-counter units.
        \label{fig-wpu-flow}}
\end{figure}

\begin{figure}
\begin{center}
\FigureFile(100mm,50mm){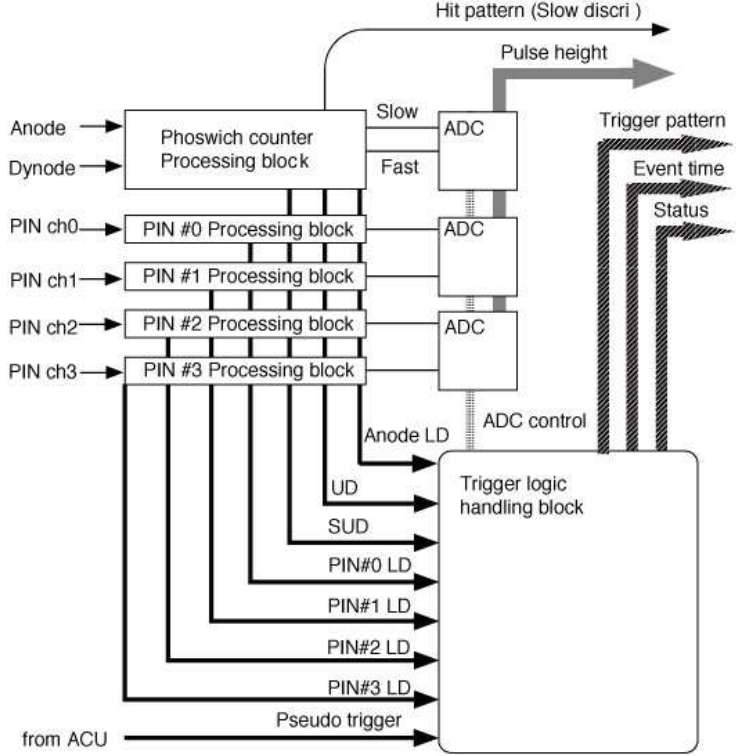}
\end{center}
\caption{Functional block diagram of the signal-processing
part  for a well counter.
        \label{fig-system}}
\end{figure}

The signal flow from a well counter unit to a WPU board and
a functional block diagram of the signal processing are
 shown in figure \ref{fig-wpu-flow} and figure \ref{fig-system}, respectively. There are two independent
circuit blocks for the PMT and the PIN.
Here we first describe the signal processing for the PMT  and
then for the PIN.

The signal from a photomultiplier anode is fed directly 
to a fast amplifier in HXD-AE to generate trigger signals.
Unlike the charge-sensitive amplifiers 
that are contained in the  HXD-S front-end electronics
and amplify dynode signals,
the anode amplifier just simply amplifies the current 
from the photomultiplier. The power consumptions
are 12.8 mW for the anode amplifier and 42.5 mW for the
dynode amplifier.
This allows us to utilize very short anode pulses, 
which are required in generating fast pre-trigger signals.
This also has a special meaning 
in the present phoswich configuration:
since the GSO scintillation has 
a faster decay time than that from BGO,
valid GSO signals produce larger anode output pulse-heights
than those from BGO,
for the same  energy deposit.
This allows HXD-AE to preferentially select  GSO signals,
while rejecting BGO events.

A preamplifier output pulse generated from a dynode signal 
is split in the HXD-AE into two shaping chains with different time constants: 
150 ns (fast shaping) and 1000 ns (slow shaping).
Each of the shaped signal is then peak held (activated by the trigger signals)
and sent to the analog to digital converters (ADCs).
Since the decay times of the GSO and BGO are different, 
the ratio of the fast and slow shaping pulse heights 
will differ between GSO and BGO.
In a fast-shaping processing chain, the charges from a BGO event are only partially integrated,
while the GSO signal is fully integrated.
In  a slow-shaping processing chain,  in contrast
almost all the charges  are  integrated for both crystals.
By comparing the fast and slow pulse heights,
we can thus discriminate GSO pulses from BGO events.
This ``double integration method'' is the basis for pulse shape discrimination (PSD).

\begin{figure}
\begin{minipage}{7.5cm}
\centerline{\FigureFile(80mm,50mm){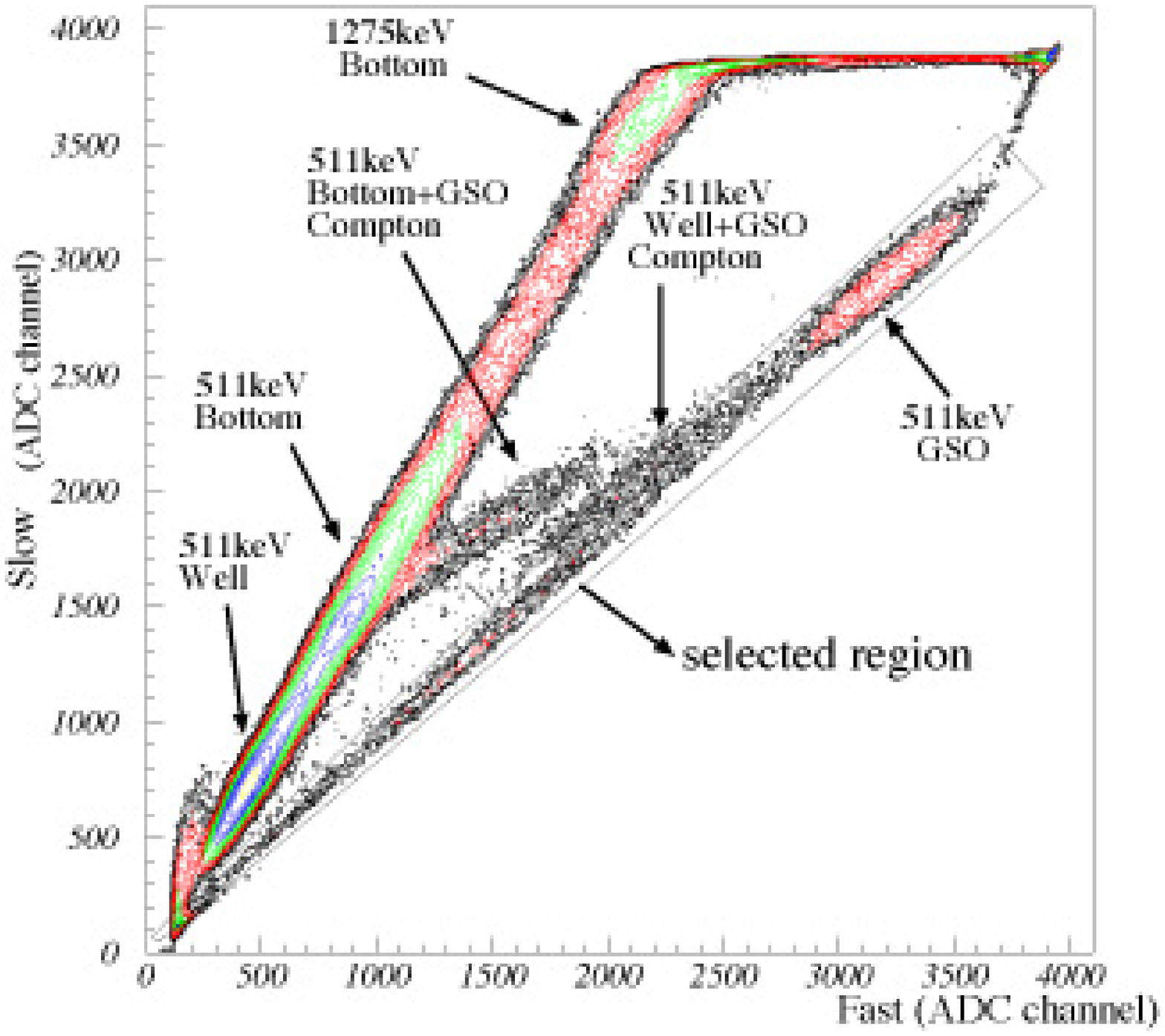}}
\end{minipage}\hfill
\begin{minipage}{7.5cm}
\centerline{\FigureFile(80mm,50mm){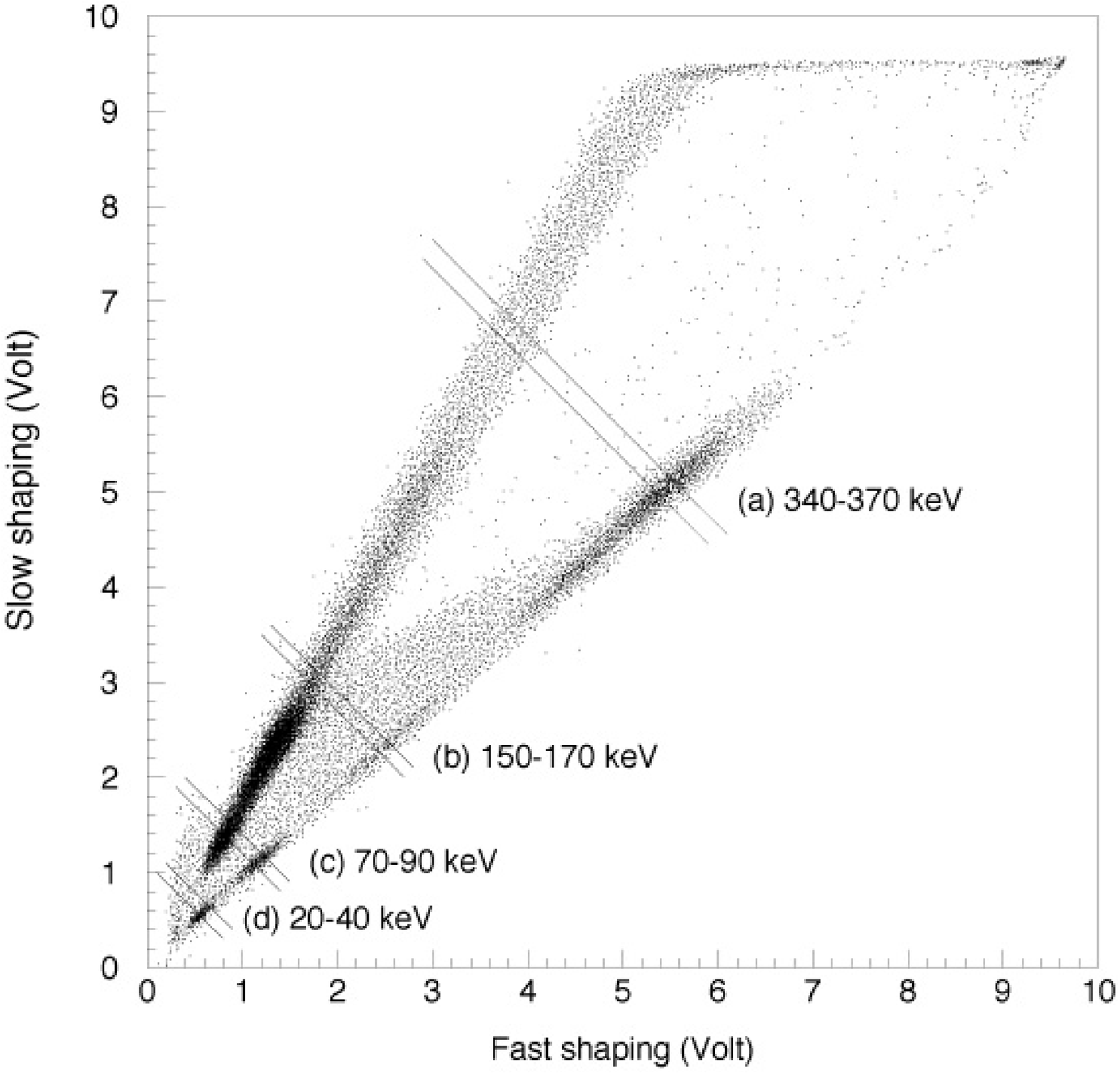}}
\end{minipage}
\begin{minipage}{7.5cm}
\centerline{\FigureFile(80mm,50mm){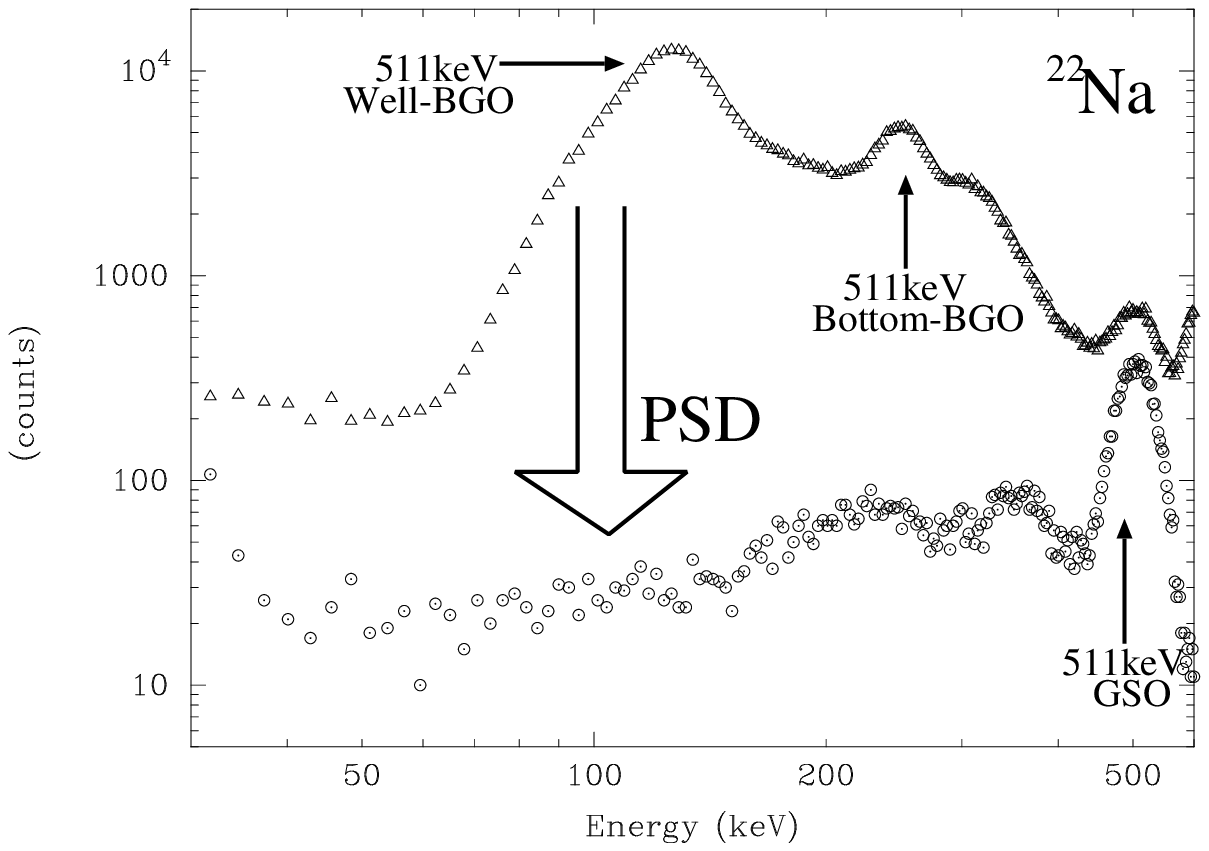}}
\end{minipage}\hfill
\begin{minipage}{7.5cm}
\hspace*{-1cm}
\centerline{\FigureFile(80mm,50mm){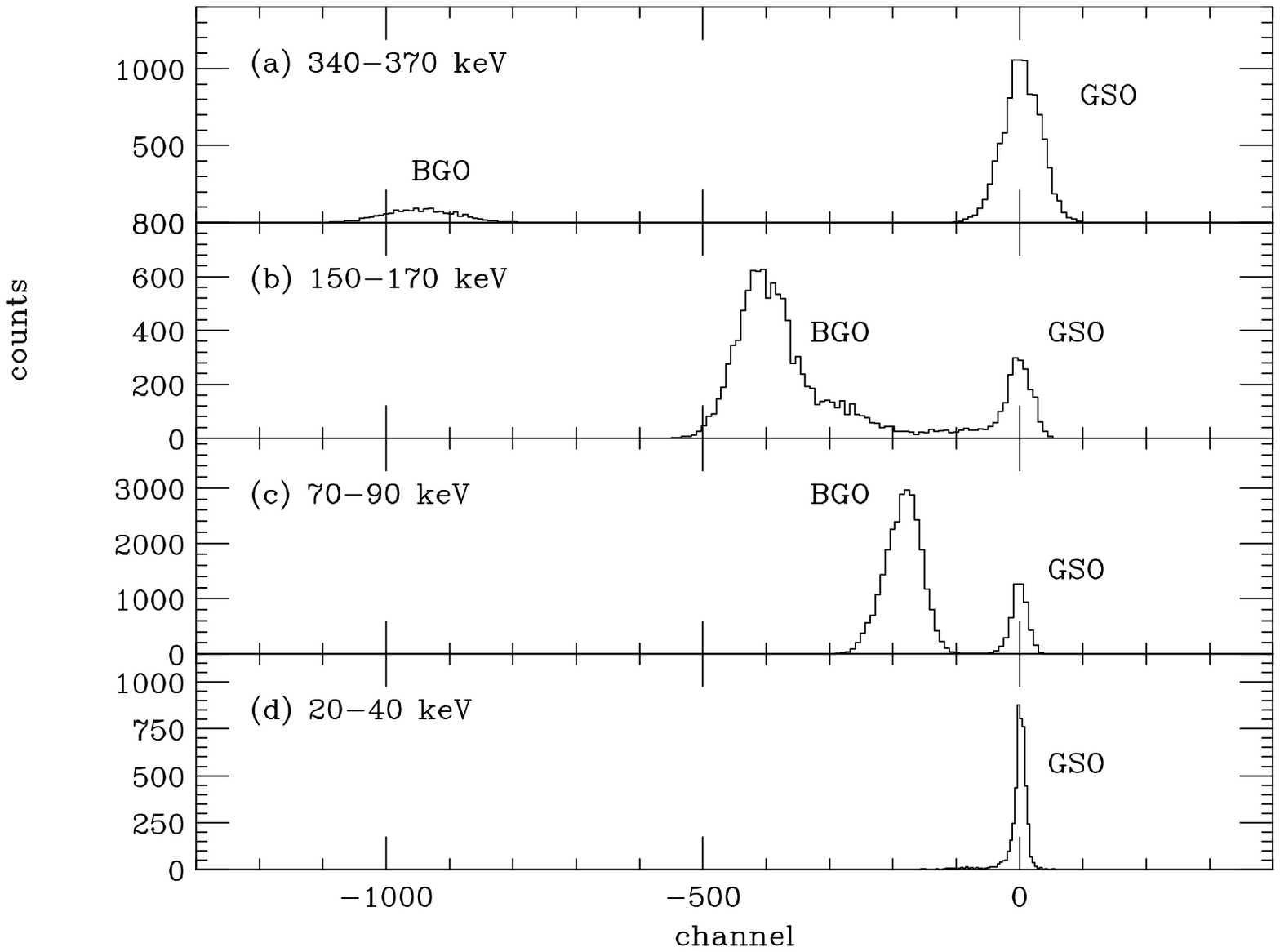}}
\end{minipage}
\caption{
(top) Fast-Slow diagram for the $^{22}$Na and $^{133}$Ba 
 irradiation, taken from a single unit.
(bottom-left) PSD selection effect on the GSO spectrum.
(bottom-right) The separation of GSO/BGO events at the lower energy. }
\label{Fig:cal-dim2}
\end{figure}

We have developed an Application Specific Integrated Circuit (ASIC) with this
PSD function \citep{Ezawa96}.
When a dynode signal is fed to this ASIC and a trigger is received  from the anode,
the ASIC gives two peak-held analog outputs,
corresponding to the two shaping time constants. The power consumption
of the ASIC is 230 mW.
These outputs can be arranged in figure \ref{Fig:cal-dim2},
called ``fast-slow diagram'' or ``two-dimensional spectrum'',
where the vertical and horizontal axes represent
the slow and fast shaping pulse heights, respectively.
In this way, we can see two lines with different inclination,
or branches,
demonstrating the separation of signals from GSO and BGO.
A bridge connecting the two branches is formed by 
Compton-scattered events,
which hit both GSO and BGO in the same unit,
because their time constants are a mixture of the two.
The actual selection is done by the ASIC in the WPU in the analog
processing chain in each channel. We can compare the two pulse heights,
either in HXD-AE, HXD-DE, or in ground analysis,
to reject  BGO events as well as Compton-scattered ones.
Since the hardware threshold
for the selection is rather loose, further selection is done by software
in HXD-DE and in ground analysis.

While the PSD technique
rejects the BGO signals in its own unit,
the HXD has another important function to reduce background,
called Hit Pattern Selection.
This function utilizes hit pattern information from multiple units,
and reject Compton scattered events that hit different units.
Information on the hits of all 36 units 
(16 well units and 20 anti-counter units) is  latched every time 
a valid event is acquired from any well counter unit.
Although all information of 36 units can be used,
the most dominant Compton scattering component
in the energy range of $\sim$600 keV
are those events scattered between the adjacent units.
Since a Hit Pattern Selection using too many detector units
will increase accidental coincidence and hence reduce the signal acceptance,
it is generally adequate to consider the 
4 adjacent units, or 8 surrounding units.
The rejection pattern can be  selected by software,
either in HXD-DE, or in the course of ground analysis.

The signals from the four PINs  inside the well-counter unit
are separately read out 
and fed individually to low-noise charge-sensitive preamplifiers (PIN preamplifiers)
mounted 
underneath the preamplifier for the PMT and fed into the circuit
in HXD-AE. Analog electronics for the PIN is designed 
 to achieve an energy threshold of 10 keV and an energy resolution 
of 3 keV (FWHM) in the orbit under power consumption of 107 mW for the
preamplifier per each PIN
diode.
 Since a PIN diode has a typical leakage current of $\sim 2$ nA at  $-20^{\rm o}$C,
and relatively large capacitance of $\sim$50 pF (including the capacitance of the
cable), we  developed a low noise preamplifier with a capacitance gradient of $\sim$15 eV/pF.
Since as
many as 64 channels of PIN diodes need to be processed,  reduction
of the size and the power consumption is very important in the AE
electronics.  We therefore integrated the last step of the amplification, the
peak-hold-circuit and the lower discriminator into the PIN
as a variant to the PSD ASIC (\cite{Ozawa97}).

\begin{figure}[h]
   \begin{center}
  \FigureFile(100mm,50mm)
{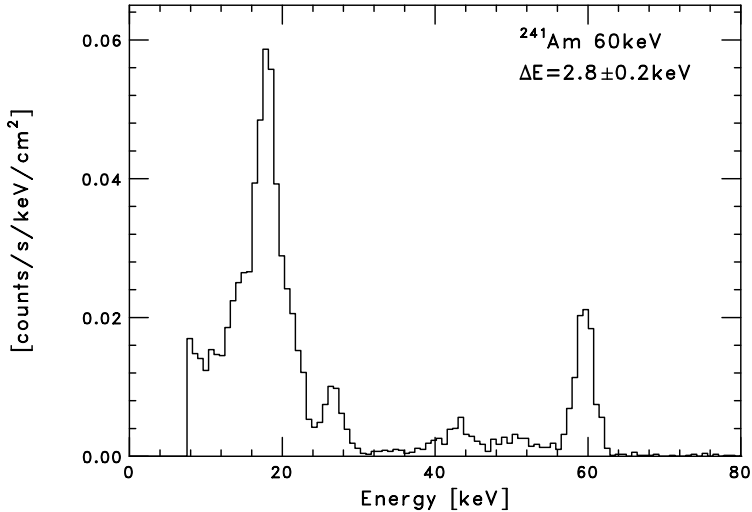}
   \end{center}
   \vspace{-0.5cm}
   \caption{PIN spectrum of $^{241}$Am obtained at  $-20^{\rm o}$C.}
   \label{Fig:Pin_Am}
   \end{figure}

By combining the preamplifier with the PIN diode
and operating them at  $-20^{\rm o}$C,
we achieved an energy resolution of 2.9 keV at 59.5 keV, 
and an energy threshold of $\sim 10$ keV,
as shown in figure~\ref{Fig:Pin_Am}.
The energy resolution of the PIN preamplifier
 is 1.0 keV in a load-free condition; this
degrades to 1.6 keV by the capacitive noise, and another 1.0 keV is due to the current noise.
The remaining  $\sim2.1$ keV is possibly caused by electronic noise in HXD-S or HXD-AE.

The trigger signal for the well-counter unit 
is generated from the anode signal, and for the PIN diodes, the output of the gain amplifier is 
fed into a comparator to generate the triggers. 
When any one of the trigger sources, including the pseudo trigger,
 is activated,  gate signals for peak holding and ADCs are generated. 
 The trigger rate in a single WPU
could  be  as high as several hundred Hz, if we turn off the
on board PSD selection. Therefore, careful design was adopted for
the digital circuits that handle the triggering.
By using the PSD information, the trigger is issued for 
events that deposited all of its energy in GSO or PIN. Signals with energies
out of the energy range of the detector is also rejected based on the upper discriminator
information. Hit-pattern information is used later in an offline analysis.
The details of trigger sequence
is described by Takahashi et al. (1998).

A significant fraction of these trigger events are charged particles, 
including  geomagnetically trapped particles 
(mainly protons) and primary cosmic rays.
Because  the energy loss of a charged particle
is proportional to the density and size of the detector, 
it becomes  very severe in the HXD,
which  employs large and heavy  scintillators
for  active shielding.
 In fact, the energy loss 
by a charged particle onto a HXD detector unit
is estimated to be typically $\sim$100 MeV, 
which is 3 to 5 orders of magnitude larger than the target energy.
Since this is well above the saturation level of the amplifying system,
a long electronic recovery time would affect
the subsequent signals,
produce spurious events, and/or increase the dead time.  
In order avoid these problems and make the system 
tolerant to these large pulses with high counting rates,
we have designed the whole electronics,
from  front-end to HXD-AE,
to quickly  recover from every energetic pulse.
The recovery time for a 100 MeV pulse is about 50 $\mu$s for
the dynode amplifier (\cite{Tanihata99}). 
Furthermore, we inhibit any subsequent  trigger
until the circuit has fully recovered from
the disturbance caused by large pulses (\cite{Tanihata99}).
The treatment for  large pulses is also implemented
in bleeder circuits in the photomultipliers  for the well-counter
units (\cite{Tanihata99}) and the anti-counter units. (\cite{Yamaoka06})

The measuring the dead time is one of the most important tasks of
the electronic system. This is
because the spectrum of a celestial object is obtained by 
subtracting the back ground spectrum from the observed one, 
and an accurate live time is especially required  when the source is faint.
In the HXD, 
a dead-time measurement is done individually for each well unit 
by counting the number of pseudo event pulse.
The pseudo event pulses are electronic pulses 
generated in the ACU module, 
periodically with a selectable period.  
These pulses are  fed into the trigger-handling
block, and activates the trigger logic in the same way as other trigger sources, 
but with a specific flag in the trigger pattern.  Since the pseudo trigger
is randomly vetoed by normal events, 
we can accurately estimate the dead-time fraction,
by comparing the total number of injected pseudo pulses
and those actually recorded in the data.
Another way to estimate the dead time is
to count up to 10 MHz (selectable by command) by  clock pulses
when the system is busy, and hence the the trigger is inhibited.  
A verification of these methods in orbit is  described in Paper II.
The dead-time counters are edited into the monitor data 
and sent to HXD-DE every second.
While the dead time counter cannot measure dead times caused
by HXD-DE capability or by telemetry limitation,
it can handle fast changes in the dead time.

The anti-counter units,
which surround the 16 well-counter units,
have two functions. 
One is to provide hit-pattern information to the well-counter units,
and the other is to monitor celestial transient events, such as
gamma-ray bursts and solar flares.
Signals from each of the 5 anti-counter units on the  same side 
of HXD-S are handled by one TPU module. 
In the TPU, hit-pattern signals from each counter are generated 
with an energy threshold of 40 keV.  
At the same time, a TPU module accumulates event
pulses from the 5 anti-counter units,
and produces their summed pulse-height spectrum,
which is sent to HXD-DE every 1 s.
If a sudden increase in the counting rate,
due, e.g., to a gamma-ray burst,  is detected,
the TPU records four-energy-band light curves
with a finer (15.625 ms or 31.25 ms) timing resolution for a certain length of time.

\subsection{Data packet from HXD-AE\label{sec:hxddataanal}}

When an event from a well-counter unit satisfies
the pre-specified trigger conditions and is hence judged to be valid,
all analog outputs from the relevant Well unit 
(those of the slow shaper, the fast shaper, and the four diode shapers) 
are digitized by ADCs and edited into  event data. 
For the signals  of the PIN diodes, the upper 8 bits are used, 
because the dynamic range of the PIN diode is small (10 to 70 keV).  
Also recorded are the occurrences of any hits in four PIN diodes and GSO 
in the well unit as a trigger pattern. The pseudo trigger and the output
from  the discriminator 
that senses overshoot of the charge-sensitive amplifier of the PIN also provides
trigger, which  are recorded in the trigger pattern.  
Any hits in the slow discriminator output from 
16 well units and 20 anti-counters are also recorded as a hit pattern. 
The output of the PSD, flags to monitor the pile-up events, 
and the upper discriminator outputs are also recorded. 
All arrival times are recorded with a resolution of $61 \mu$s 
during the normal operation mode. 
The data, formatted as shown in  table \ref{Table:event-data}, are 
 sent to the HXD-DE on an event-by-event 
basis.  
Thus, one valid event occupies 18 Bytes.
In addition to the well-event data,  
a number of additional data in the HXD system used to monitor  its proper 
function and to estimate of the background and dead time are collected.

\begin{table}
\caption{WPU Envet Data}
\begin{center}
\begin{tabular}{|ll|}
\hline
Channel ID & 2 bit \\
Event Time&  19 bit\\
Phoswich Counter Slow Pulse Height & 12 bit\\
Phoswich Counter Fast Pulse Height & 12 bit\\
PIN detector 0 Pulse Height & 8 bit \\
PIN detector 1 Pulse Height & 8 bit \\
PIN detector 2 Pulse Height & 8 bit \\
PIN detector 3 Pulse Height & 8 bit \\
Trigger Pattern & 7 bit \\
PIN LD Trig & 1 bit \\
Hit Pattern & 36 bit \\
PSD Out & 1 bit \\
PMT UD & 1 bit \\
PIN UD & 1 bit \\
PMT Double Trig Flag & 1 bit \\
PIN Double Trig Flag & 1 bit \\
Reset Flag  & 1 bit\\
\hline
\end{tabular}
\end{center}
\label{Table:event-data}
\end{table}

\subsection{Digital Electronics (HXD-DE)}

HXD-DE is the CPU-based signal processing part of the HXD.
It is designed:
(1) to provide the primary interface with the satellite data processor 
for command and telemetry, 
(2) to control the acquisition and formatting of the data from HXD-AE, 
and (3) to react to events such as $\gamma$-ray bursts. 
HXD-DE is based on a CPU 80C386 running at 12 MHz.
It has two CPU boards, with one board working
while the other is idling for backup purposes.
Mounted on the CPU  board is 512 KB of SRAM 
used for a programmable  area, 
another 512 KB of SRAM serving as direct memory access  (DMA) buffer, 
and 512 KB of EEPROM for  program storage.

One of the baseline requirements of HXD-DE  is 
to acquire data without loss from sources 
that are several times as bright as the  Crab Nebula. 
This means that HXD-DE should be capable of 
an acquisition rate of at least several hundred
events per second. If we loosen the electronics parameters
such as the level of PSD discrimination and low-level
discriminator, the rate goes up quickly.
Therefore we set our rate tolerance of system to be
4,000 events per second. In order to achieve this high acquisition rate with limited hardware resources, 
we use a fast real time operating system
specially designed for this experiment, 
in conjunction with a circuit capable of receiving data  from HXD-AE 
by the DMA mode.
When data are sent or received in the DMA mode, 
the I/O adaptor accesses the DMA buffer directly, 
without interrupting the CPU.

In orbit, irradiation by charged particles can
cause bit errors in the memory.
In order to prevent these errors from causing malfunctioning,
the process called Memory-Patrol is always running to
verify the contents of memory byte by byte. 
In addition, the HXD-DE data are accompanied by 
an error correction code,
which repairs 1-bit errors,
and reports the occurrence of 2-bit (or more) errors.

\subsection{PI PROGRAM}

The process called ``PI program'' is an important addition 
to the  HXD-DE software, and is used to realize high signal throughput
while keeping a low background capability. 
{Its task, for example, is}
to filter out those background events that 
leak through the hardware selection in WPU, 
and to compress the ``transient'' data sent from the TPU. 
These functions becomes particularly important 
when the telemetry bandwidth is limited.
Additional tasks of the PI program are to collect calibration data, 
and to notify $\gamma$-ray burst triggers   to the TPU modules.
The PI program is activated when the data arrives from the DMA buffer
with the data type as input. 
According to the data type, 
the main routine of the PI program calls 
the corresponding function that handles the data.

Although the hardware PSD in the WPU module is very powerful, 
the PI program provides more  flexible ways 
to further select events.
The PI program does the event selection
via several different algorithms,
which can be chosen by commands. 
Several examples of these algorithms are given below:

\begin{description}
\item  [hit pattern:]
The program performs event selection based on the
hit pattern information  of the surrounding units.
Using this, we can reject Compton scattered events 
and particle-interaction events. 
\item [pulse shape:]
In HXD-AE,  pulse-shape discrimination is performed by comparing
the pulse heights between  the slow and fast shaping amplifier outputs.
In HXD-DE, finer and more flexible selections
can be applied on the fast-slow diagram.

\item [delta-t:]
In some inorganic scintillators, such as CsI(Tl), 
a train of pulses with random shapes occur
during 400 $\mu$s - a few ms after a large energy deposit
(at least E$\geq 60$ MeV) in the crystal  (\cite{Takahashi93}). 
This leads to a number of false triggers in a very short time. 
These false  events can be removed in the HXD-DE,
because the PI program can measure the
time difference between successive events 
and remove them if the difference is shorter than a certain threshold.
\end{description}

\section{Ground Calibration and Tests}

Before the launch, the HXD performance
was studied extensively on the ground.
Since the HXD is a highly sophisticated experiment
involving a large number of components,
a comprehensive data base has to be prepared 
before launch in order to construct a set of software 
required to deduce scientific results from the observational data.
In  ground pre-launch tests,
we first qualified the basic system functions,
including the pulse shape discrimination, 
hit pattern generation, 
and the methods of dead time measurements.
Subsequently,  
we placed HXD-S in a low-temperature environment ( $-20^{\rm o}$C),
irradiated it with various gamma-ray  emitting radio isotopes,
and acquired data via HXD-AE and HXD-DE.
We thus measured the basic characteristics
of individual well-counter and anti-counter units, 
including the gain linearity and energy resolution,
as well as their angular responses.
The background spectrum was also  measured 
to verify the low background performance of the HXD.

\subsection{Alignment}

\begin{figure}
\begin{center}
\FigureFile(80mm,50mm){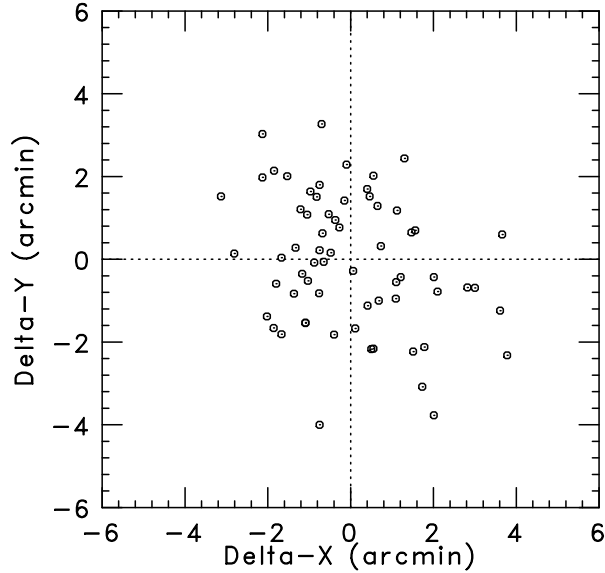}
\end{center}
\caption{Pointing offsets of the fine collimators in $X$- and $Y$- directions
measured by scanning a radio active source in front of HXD-S.}
       \label{Fig:Align}
       	\end{figure}

The alignment of the 64 fine collimators has been measured by means 
of optical laser light and a $\gamma$-ray source. 
Figure \ref{Fig:Align} shows their angular offsets 
measured by scanning a $^{133}$Ba  
source in front of HXD-S at some distances.
These two independent measurements  confirm that
the collimators are aligned  with an accuracy of $3.5'$ (FWHM). 
This ensures an effective transparency of 90\%,
when a target is placed at the mean direction 
of the optical axes of the 64 fine collimators. 
We verify that the angular offsets did 
not change after the vibration test of HXD-S.
The effective area shown in figure \ref{Fig:hxd-eff} has to be collected
with this effective transparency. The overall calibration for the absolute
normalization in orbit is described in Paper II.

\subsection{Spectral performance}

Calibrations of gains and offsets and measurements of energy resolution 
are essential in constructing the instrumental response function,
which in turn is needed to reconstruct  incident spectra from
the observed pulse-height distributions. 
Before assembling HXD-AE,
we measured  the linearity and  dynamic range of its 
individual signal-processing circuits using electronic test pulses. 
After some iterative electronics gain and offset adjustments, 
the relations between the test pulse amplitude  
and the resultant ADC values of individual signal channels
were measured,
and were registered as  gain parameters in a calibration data base.

When mono-energetic gamma-rays 
are incident on the PIN or GSO detector,
the output pulse height distribution 
primarily consists of a roughly Gaussian peak 
formed by full-energy-deposit events.
Although the Gaussian centroid  is roughly proportional to the gamma-ray energy,
we must consider the measured electronics non-linearity (see  above).
Furthermore,
GSO scintillator is known to exhibit some non-linearity 
between its energy  deposit and  light output,
particularly toward lower energies (\cite{Uchiyama01});
this has to be calibrated.
The pulse-height distributions also 
very often show secondary  features with lower pulse heights. 
In the GSO spectra, the strongest secondary feature
is produced by  gamma-rays that undergo 
Compton scattering in the detector and escape from it. 
In the PIN spectra, guard-ring structures in the diode produce 
a spectral sub-peak with a $\sim 2/3$ of  the main-peak
pulse height \citep{Sugiho01}.
Since these secondary features generally depend on the
detector geometry,
we use Monte Carlo simulation code, either  EGS4 (\cite{Nelson85}) or Geant4 (\cite{Allison2006,Agostinelli2003}),
to compute the response of the HXD detector
as a continuous function of energy (\cite{Terada04,Kokubun2006}). 
These codes allow us to handle the detailed detector geometry, such as the guard ring and housing of the PIN diodes. 
These Monte Carlo simulations refer to the data base,
which summarizes parameters of individual GSO and PIN detectors
such as the linearity  and  the energy resolution,
measured at discrete energies using isotopes by fitting the spectrum 
with a Gaussian profile. Sub peak structures seen in the PIN spectrum are
properly taken into account in the code.

\begin{figure}
\centerline{\FigureFile(100mm,50mm){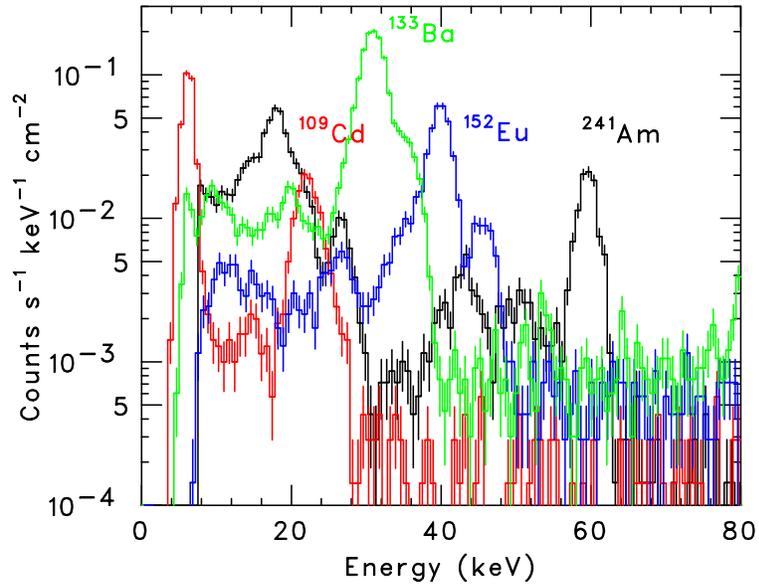}}
\caption{ Composite energy spectra  of the PIN diode,
 from various  $\gamma$-ray 
isotopes, are superposed for presentation (W01 for $^{241}$Am, W12 for $^{109}$Cd,
W23 for $^{152}$Eu, and W33 for $^{133}$Ba.}
  \label{Fig:PIN-spec-line}
\end{figure}

\begin{figure}
\begin{center}
\FigureFile(100mm,50mm){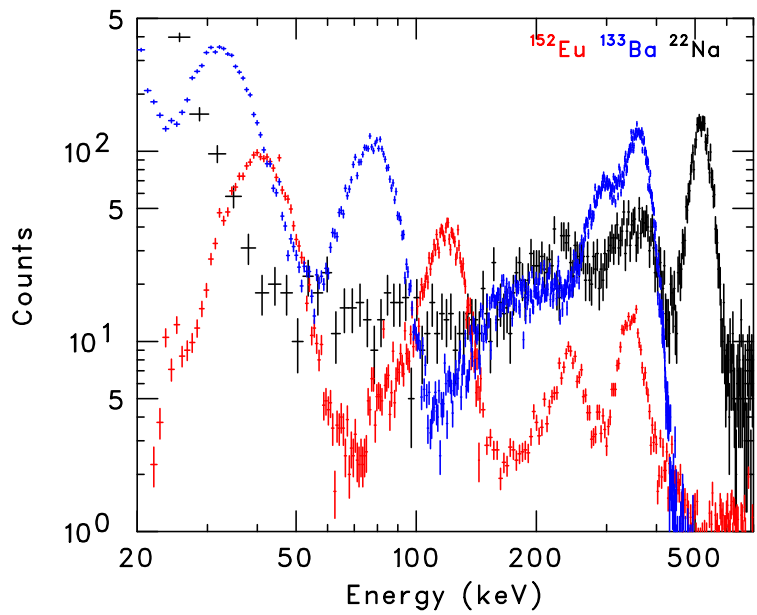}
\end{center}
\caption{The same as figure~\ref{Fig:PIN-spec-line}, but for GSO.}
        \label{Fig:GSO-spec-line}
	\end{figure}

\begin{figure}
\begin{center}
\FigureFile(80mm,50mm){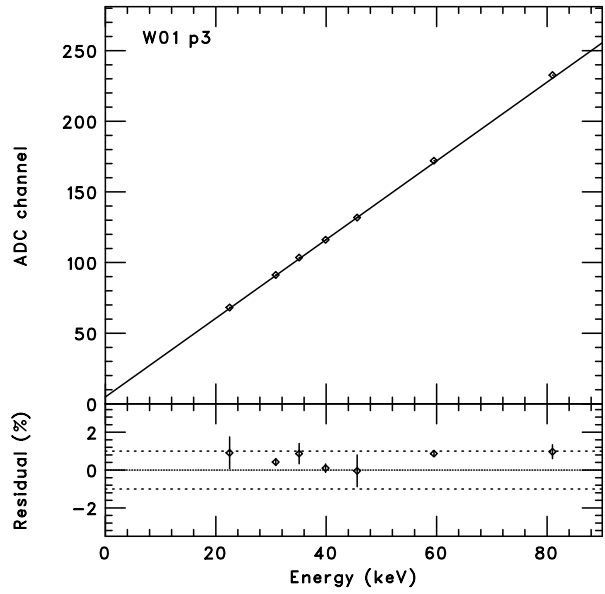}
\end{center}
\caption{Pulse-height vs energy linearity of   the well-counter unit W01 (PIN03).}
       \label{Fig:PIN_lin}
       	\end{figure}

\begin{figure}
\begin{center}
\FigureFile(80mm,50mm){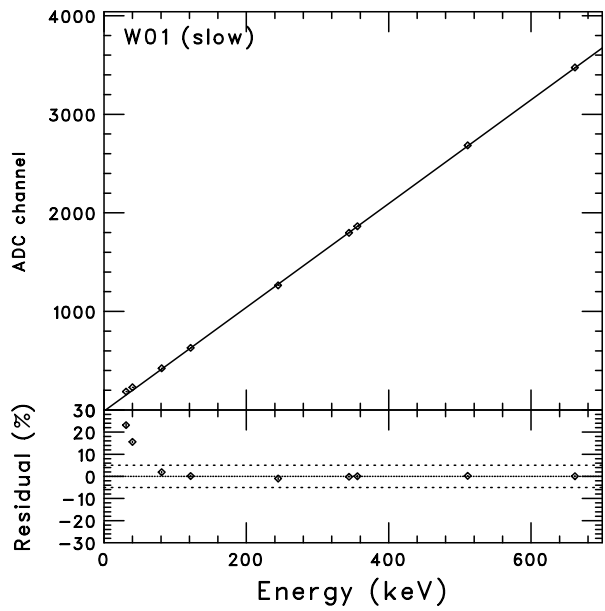}
\end{center}
       \caption{The same as figure~\ref{Fig:PIN_lin}, but for GSO in the
 well-counter unit W01.}
       \label{Fig:GSO_lin}
	\end{figure}

After  HXD-S and HXD-AE were assembled,
we measured the  linearity and energy resolution,
including both the sensor and electronics properties, utilizing 
gamma-ray lines from radio 
isotopes listed in Table \ref{Table:tab-fm-res}.
The obtained PIN and GSO spectra are 
shown in figure \ref{Fig:PIN-spec-line}
and figure \ref{Fig:GSO-spec-line}, respectively.
The GSO spectrum clearly reveals
the 31 keV K$\alpha$ line from  $^{133}$Ba,
indicating that the lower threshold energy of GSO has been made sufficiently low.
The phoswich configuration successfully prevents 
the Compton events from contaminating the GSO spectra,
and makes the full-energy peaks much more prominent
than in the spectra taken with conventional scintillators.
Figure \ref{Fig:PIN_lin} and  figure~\ref{Fig:GSO_lin} 
shows  linearity plot of one representative PIN  and  GSO, 
respectively, obtained from these $\gamma$-ray line measurements.
Below $\sim 100$ keV, 
the GSO results show the slight non-linearity mentioned above,
which can be corrected by the analysis software.

	\begin{figure}
\begin{center}
\FigureFile(80mm,50mm){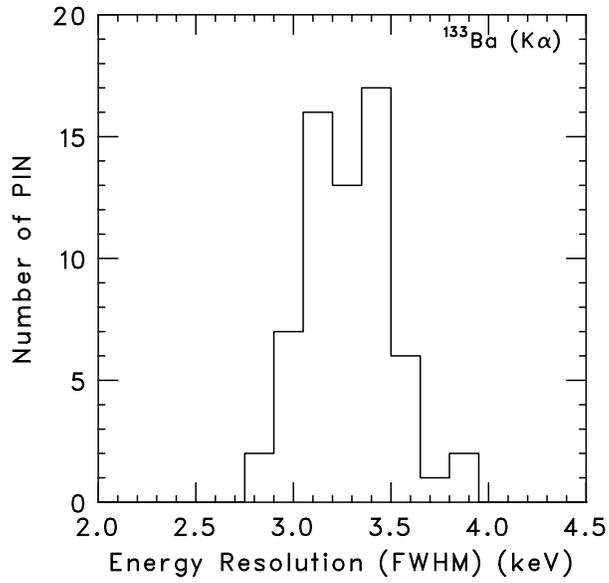}
\end{center}
\caption{Distribution of the energy resolution among the 64 PIN diodes in the HXD,  
measured at  $-20^{\rm o}$C using the 31.5 keV $\gamma$-ray line from $^{133}$Ba .}
\label{Fig:dist_res_pin} 
	\end{figure}
		
	\begin{figure}
\begin{center}
\FigureFile(80mm,50mm){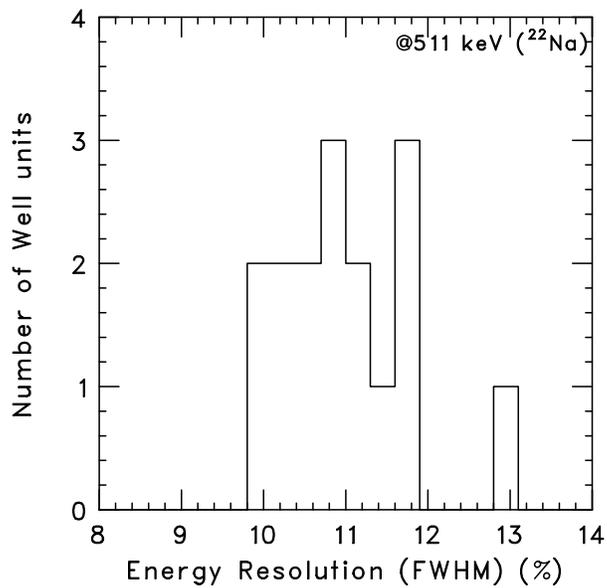}
\end{center}
\caption{
Distribution of the energy resolution of GSO  in the 
16 well-counter units, measured using the 511 keV 
$\gamma$-ray line from $^{22}$Na.}
\label{Fig:dist_res_GSO} 
	\end{figure}

	\begin{figure}
\begin{center}
\FigureFile(80mm,50mm){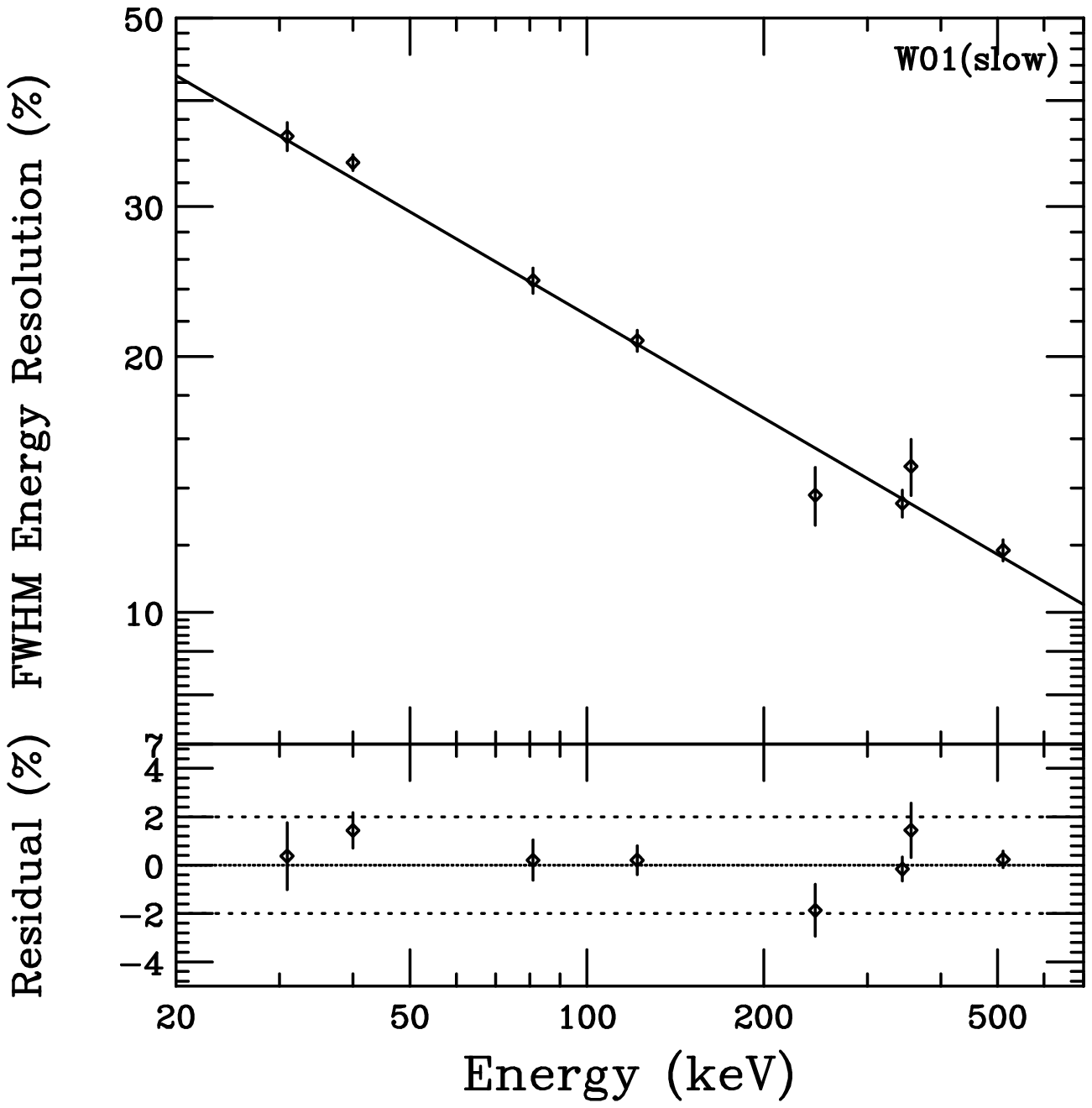}
\end{center}
\caption{Energy dependence of the GSO energy resolution measured with the unit W01.}

       \label{Fig:GSO_res}
	\end{figure}
	
	As can be inferred from figure~\ref{Fig:PIN-spec-line},
and summarized in figure \ref{Fig:dist_res_pin},
the typical energy resolution of the PIN diodes has been
obtained as $\sim 3.3$ keV  at  $-20^{\rm o}$C.
This meets our design goal, 
because then the low-energy threshold of PIN can be
lowered to $\sim 10$ keV,
ensuring an overlap with the energy range 
of the XIS \citep{Koyama2006}. 
Similarly, figure~\ref{Fig:dist_res_GSO} shows the
distribution of energy resolution of the GSO scintillator
in the 16 well-counter units,
while figure~\ref{Fig:GSO_res} shows the energy resolution of unit W01 as a function 
of energy.
Thus, the energy resolution is 12\% at 511 keV
(or 10\% at 662 keV), 
which again  meets our design goal.
The corresponding value of a single GSO of 5 mm thick 
read directly by a PMT is typically 7\% at 662 keV,
and its degradation to $\sim 10$\% is caused by the phoswich configuration:
about 30\% of the GSO scintillation photons are absorbed 
when they penetrate the BGO bottom piece, 
and another $\sim 20$\% is reflected into the BGO top piece.

\begin{table}[htbp]
\caption{Radio isotopes used for the calibration of the PIN and the GSO detectors in the
Well-counter units. }
\label{Table:tab-fm-res}
\begin{center}
\begin{tabular}{|c|c|c|}
\hline \hline
Isotope & Energy (keV) & \\
\hline
$^{109}$Cd &  22  & PIN    \\
$^{133}$Ba  & 31.5  & PIN\\
$^{152}$Eu & 41.1  & PIN / GSO\\
$^{241}$Am & 59.5  &  PIN / GSO\\
$^{133}$Ba & 81    &  GSO\\
$^{152}$Eu & 122   & GSO\\
$^{152}$Eu &  344   & GSO\\
$^{22}$Na & 511   & GSO\\
\hline
\end{tabular}
\end{center}
\end{table}

\subsection{Background level and expected sensitivity}

\begin{figure}
\begin{minipage}{7.5cm}
\centerline{\FigureFile(80mm,50mm){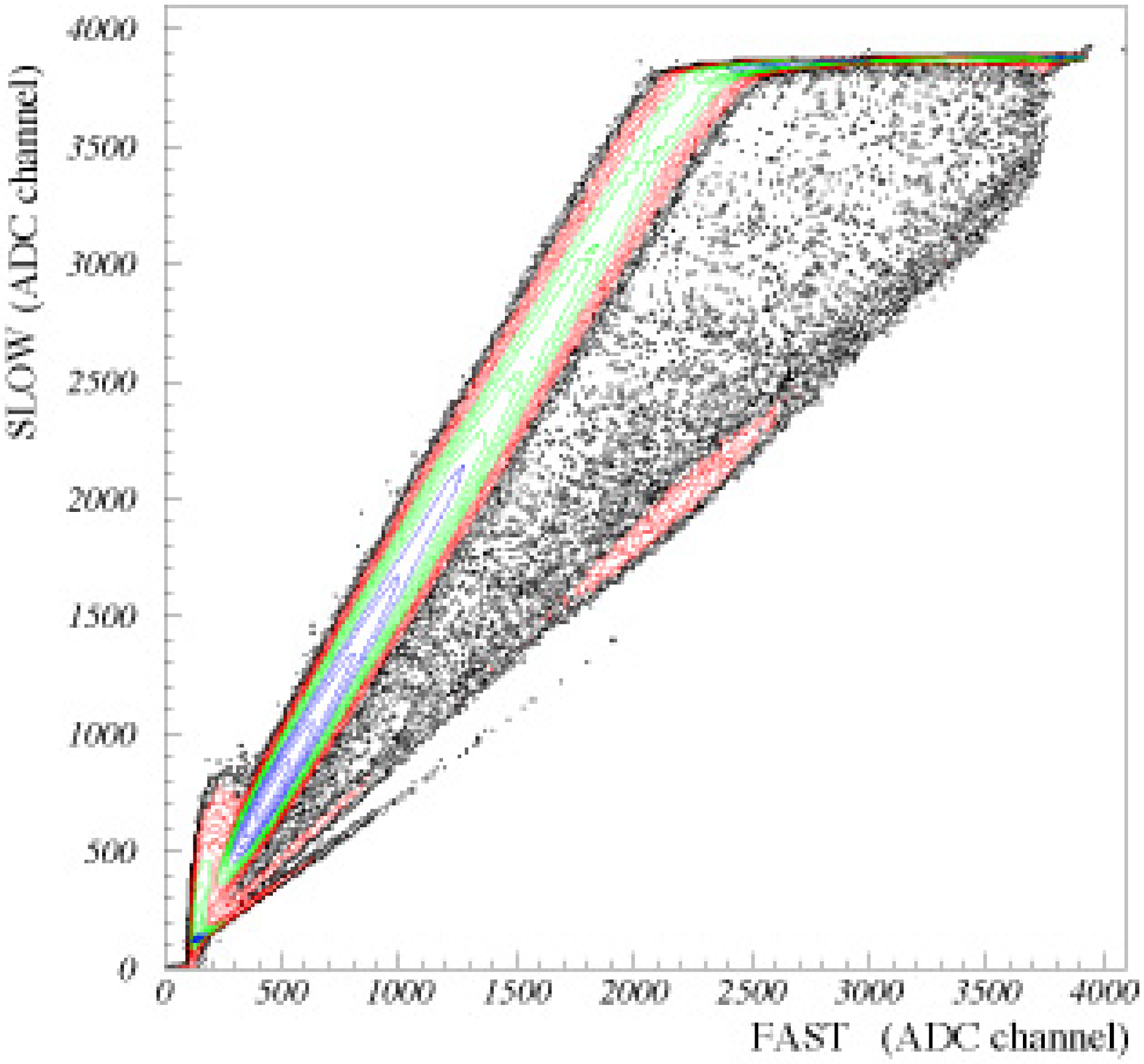}}
\end{minipage}\hfill
\begin{minipage}{7.5cm}
\hspace*{-1cm}
\centerline{\FigureFile(80mm,50mm){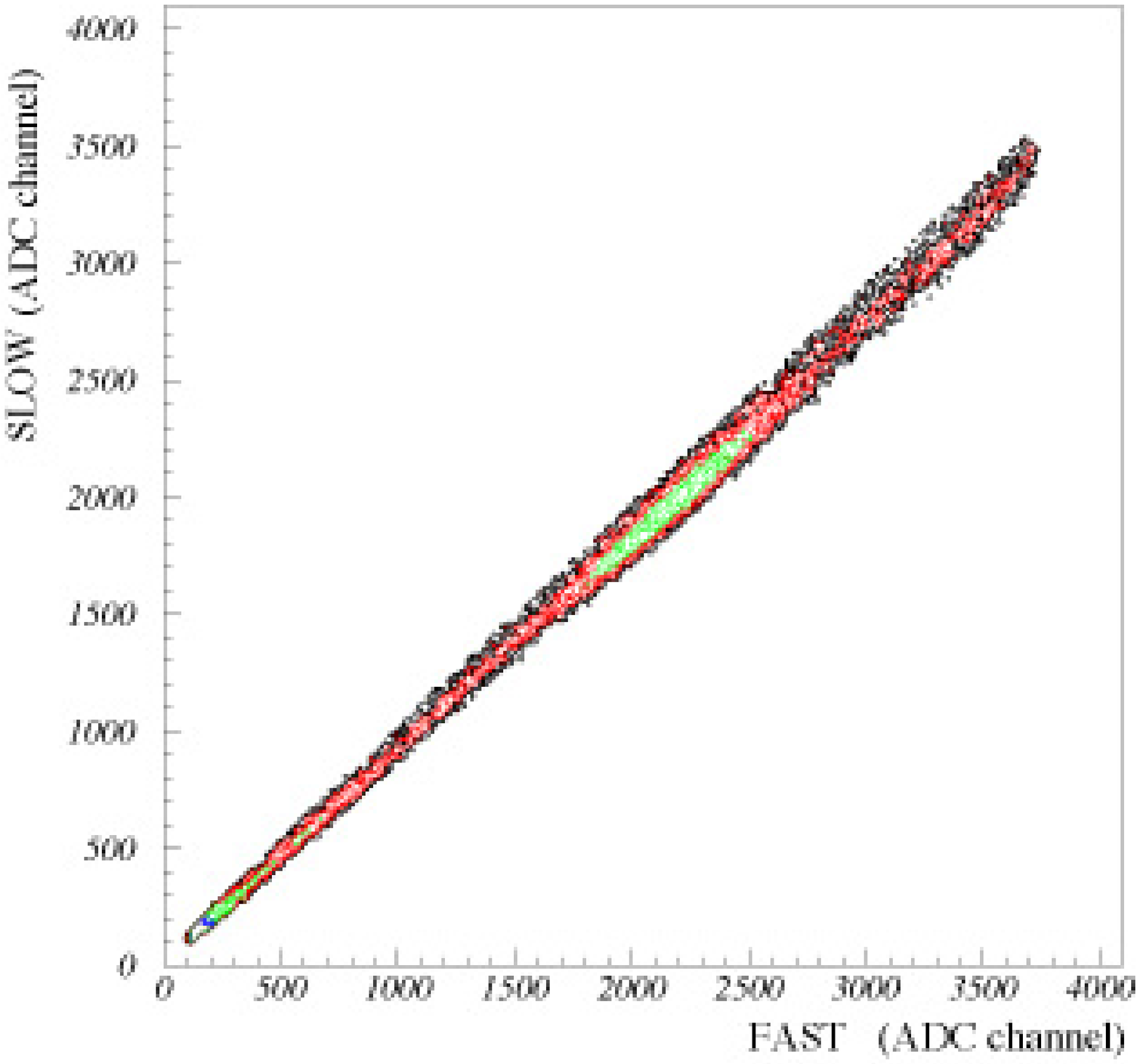}}
\end{minipage}
\caption{Background spectrum  of a typical well-counter unit measured
on the ground, presented on  a fast-slow plane.        
(a) Total  background events detected by the phoswich counter.
(b) Events selected by the PSD (see text) , both taken from a single
unit }
\label{Fig:cal-bgd}
\end{figure}

It is of vital importance to confirm
that the HXD achieves a sufficiently low  background level,
after the event selections implemented in the HXD are applied. 
Figure \ref{Fig:cal-bgd}(a) shows 
a ground background spectrum of a typical well-counter unit,
acquired in these pre-launch tests.
It is presented in the same fast-slow diagram as figure \ref{Fig:cal-dim2}.
Again, we clearly observe the separation of 
the GSO and BGO branches.
Below the GSO branch,
we can see a narrow line.
Because of the very fast time constant attributed to these events,
they are thought to be due to Cherenkov light
produced in PMT glass.
Figure \ref{Fig:cal-bgd}(b) shows the events selected by the PSD
in the data analysis;
the selection  here is somewhat tighter 
than is implemented in HXD-AE as hard-wired function (subsection 4.1),
but comparable to those that will be employed in actual observations.
It is apparent that all of the BGO and Compton events
are effectively rejected by the PSD cut. Further details on the selection 
criteria in the fast-slow plane are described in Tanihata et al. (1998) and Paper II.

\begin{figure}
\centerline{\FigureFile(100mm,50mm){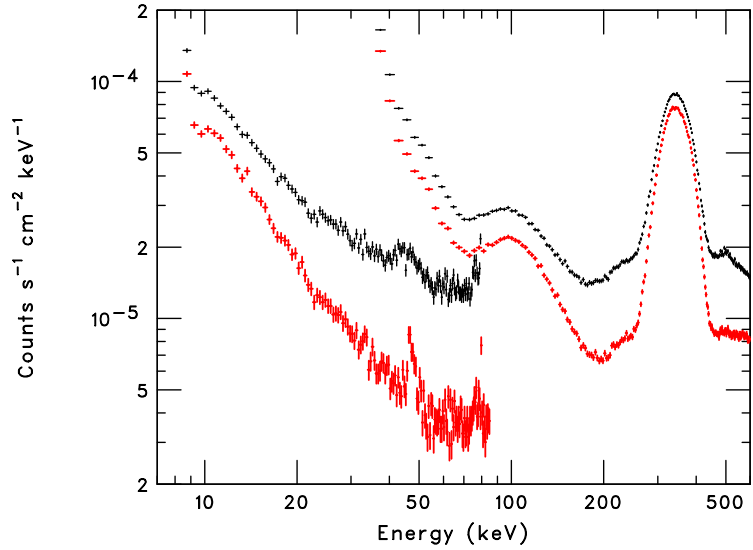}}
\caption{Background spectrum obtained from a ground calibration.
	The black line represents the spectrum selected by 
	PSD, and the red line represents the spectrum
	additionally selected using the hit pattern
	from  4 adjacent units for rejection.
        \label{Fig:background_spec}}
\end{figure}

Figure \ref{Fig:background_spec} is 
the PIN and GSO background spectra thus obtained 
on ground after the PSD selection;
the latter spectrum is  
equivalent to the horizontal projection of the 2-dimensional
spectrum in figure \ref{Fig:cal-bgd}(b).
Thus, the PIN background spectrum is rather featureless
except  Gd K lines from the GSO scintillators seen at $\sim 45$ keV,
while that of GSO exhibits a prominent peak around 360 keV.
At this stage, the measured on-ground background level thus 
becomes $\lesssim 1 \times 10^{-4}$ cnts s$^{-1}$ cm$^{-2}$ keV$^{-1}$
over the entire HXD energy range, 10--600 keV.

The remaining  background  events are further 
reduced by taking anti coincidences among the 16 well-counter
and 20 anti-counter units, employing the hit pattern information
attached to individual GSO events (subsection~\ref{sec:hxddataanal}).
The solid line in figure \ref{Fig:background_spec}  shows the
background spectrum after   hit pattern selection
by adjacent 4 units  for
both GSO and the PIN diodes. 
For the PIN diodes, the events 
that hit the BGO or GSO in its own well unit are also rejected.  
The effect of anti-coincidence among units by utilizing hit pattern information 
reduces the background level by 30--50 \%.
In the GSO selected spectrum (figure \ref{Fig:background_spec}),
it is apparent that 
while the background level under 100 keV is determined by 
off-aperture X-rays from the FOV, 
it is dominated by the intrinsic radio activation above 100 keV.
A peak around 360 keV is clearly seen.
This is due to  the intrinsic radioactivity of the 
$^{152}$Gd in the GSO crystal 
(abundance 0.2\%, 2.14 MeV $\alpha$-ray,
 half-life 1.1 $\times$ 10$^{14}$ year) (\cite{kamae1993}).
The number of photons of the peak is counted to be 
0.2 cts/s per well, which is consistent with the
calculated value.  
Though this intrinsic peak cannot be removed, we
utilize it as an in-orbit calibration source, while we use activation lines of 
GSO and BGO for PIN (Paper II).
The resulting background level at the sea level was
$\sim$ 1$\times$10$^{-5}$ cts s$^{-1}$ cm$^{-2}$ keV$^{-1}$ at 30keV 
for the PIN diodes, and 
$\sim$ 2$\times$10$^{-5}$ cts s$^{-1}$ cm$^{-2}$ keV$^{-1}$ at 100 keV,
and $\sim$ 7$\times$10$^{-6}$ cts s$^{-1}$ cm$^{-2}$ keV$^{-1}$ at 200 keV for the
phoswich counter.  
This is the lowest background level we have reached
throughout our development, 
including the balloon experiments (\cite{Takahashi93}).

\begin{figure}
\centering
\FigureFile(100mm,50mm){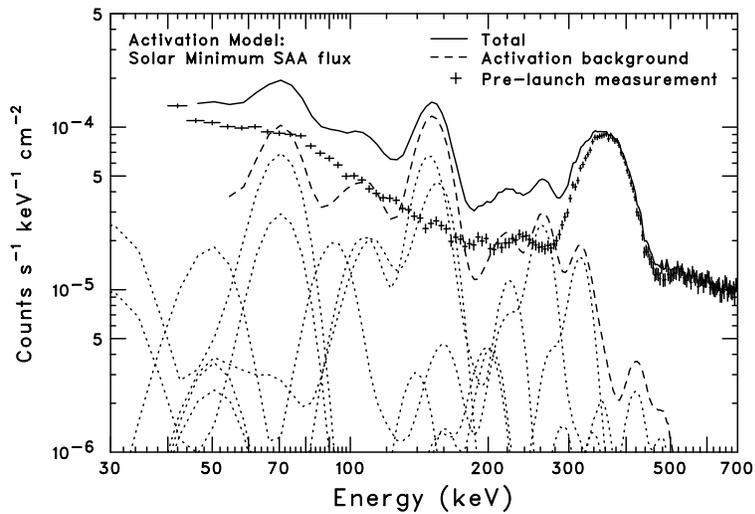}
\caption{Expected in-orbit GSO background spectrum  
at energies  above 40 keV (solid line).
It is based on the measured on-ground background (crosses),
and includes estimated activation backgrounds 
(dashed line)
calculated from our beam experiment \citep{Kokubun99}.
The dotted lines show individual activation components.
}
\label{Fig:ExpectedBack}
\end{figure}

In orbit, the background level is expected to increase
because charged particles penetrating through the scintillators
will produce various unstable nuclear isotopes.
We have thus carried out proton irradiation experiments 
using particle accelerator facilities, 
in which a flight-equivalent well and anti units were both irradiated by protons 
accelerated up to typical energies (100--200 MeV)
 of the geomagnetically trapped cosmic-rays in orbit.
Figure \ref{Fig:ExpectedBack} represents the expected background spectrum in orbit,
obtained by adding together the measured on-ground background
and the expected increase due to radio-activation.
The latter was calculated assuming a proton flux 
in an orbit of altitude of 550 km and inclination of 31 $^\circ$at the solar maximum.
 \citep{Kokubun99}.

The sensitivity of a detector for a weak continuum source is generally 
 determined by statistical and systematic uncertainties in the background,
the effective area   in the relevant energy range, and the observation time.
The  very low background level  demonstrated
by the ground calibration ensures  a high sensitivity 
 for the HXD,
even with its  relatively small effective area.
 The expected sensitivity will be presented in the companion paper (Paper II),
taking into account the actually measured in-orbit background and its variation. 

In order to achieve a high sensitivity, precise modeling of the background
is important. For this purpose, we turn on the PIN diodes even in the SAA passage
to count the number of energetic charged particles  that have passed through PIN
diodes. By utilizing the count rate of upper discriminators attached to each
analog processing chain for the PIN diode, the information 
about the nature of non X-ray background can be obtained by measuring
a variation and distribution of such high energy particles. Detail results
of measurements and their application to the modeling are described 
in Paper II.

\section{Summary}

The HXD has been designed to
achieve an extremely low in-orbit background ($\sim 10^{\rm -4}$~c
~s$^{\rm -1}$~cm$^{\rm -2}$~keV$^{\rm -1}$), based on a combination of novel
techniques: (1) five-sided tight BGO  active shielding;
(2) the use of the 20 shielding counters made of thick BGO crystals
that surround the 16 main GSO/BGO counters; (3) sophisticated onboard
signal processing and onboard event selection, employing both
high-speed parallel hardware circuits in the Analog Electronics, and
CPU-based signal handling in the Digital Electronics; and (4) a
careful choice of materials that do not become strongly radio-activated
under in-orbit particle bombardment.  
Finally, (5) the narrow field of view below $\sim 100$ keV, defined by the fine collimator,
effectively reduces both the CXB contribution and any source
confusion.  
A detailed pre-flight calibration confirms that the performance of the HXD meets
the design goal of the experiment,
including the threshold of $\sim 10$ keV of the PIN diode,
and the very low on-ground background of 1 -- 5 $\times 10^{-5}$
counts sec$^{-1}$ keV $^{-1}$ cm $^{-2}$.

The authors thank  former graduate students : 
Hajime Ezawa, Keiichi Matsuzaki, 
Eriko Idesawa, Mutsumi Sugizaki,
Satoko Osone, Ginga Kawaguchi, Kyoko Takizawa,  and  Toshihisa Onishi
 for their help to the HXD project. 
We wish to thank many people involved in the development of the HXD; 
Hitachi Chemical, Shin-Etsu Chemical, Hamamatsu Photonics, 
Baikowski Japan, Super-Resin Industry, Takachiho Seisakusho, and
C.I. Industry, in particular, to  K. Masukawa, T. Itoh, H. Ishibashi, K. Yamamoto, and A. Okada
 for their assistance throughout the project and
Isao Odagi, Yasuhisa Tanaka, Keiji Sato, and  Noboru Morita
for  their  contribution to the HXD-DE for  the  Astro-E satellite.

\section*{Present Address}

\begin{list}{}{\setlength{\leftmargin}{0.5 cm}
        \setlength{\itemindent}{-0.5 cm}
        \setlength{\parsep}{0.0 cm}
        \setlength{\itemsep}{0.0 cm}}

\item $^\ast$ Fujitsu System Solutions Ltd.,
 	Bunkyo Green Court Center Office, 2-28-8 Honkomagome, Bunkyo-ku,
 	113-0021
\item $^\dag$  Mitsubishi Electric Company,325 Kamimachiya,
Kamakura,247-8520
\item $^\ddag$ Nihon University, 7-24-1, Narashinodai, Funabashi, 274-8501
\item $^\S$ Goldman Sachs, Japan Ltd. 6-10-1, Roppongi, Minato-ku, 101-6144
\item  $^\P$  NASA/Goddard Space Flight Center, Greenbelt, MD 20771, USA.
\item  $^|$  Ministry of Economy, Trade, and Industry, Kasumigaseki,
Chiyoda, 100-8917
\item $^{\ast\ast}$ Tokyo Institute of Technology, 2-12-1 Ookayama,
Meguro, 152-8551
\item   $^{\dag\dag}$Japan Patent Office,   Kasumigaseki, Chiyoda, 100-8915
\item $^{\ddag\ddag}$ deceased
\item $^{\S\S}$  Tecnets Co., 3-3-5 Owa, Suwa, 392-8502
\item $^{\P\P}$  Agilent Technologies Japan, Ltd.,
  		9-1 Takakura-cho, Hachioji, 192-8510
\item $^{||}$ NEC-Toshiba Sapce Systems, Ltd.,
  		1-10 Nissin, Fuchu,183-8551
\item $^{\ast\ast\ast}$ ITFOR Inc, 21 Ichibancho, Chiyoda-ku, 102-0082
\item $^{\dag\dag\dag}$ Hitachi Ltd., 7-1-1 Omika-machi, Hitachi, 319-1292
\end{list}

\end{document}